\documentclass[reprint,notitlepage,showpacs,preprintnumbers,nofootinbib,amsmath,amssymb,aps,prd,floatfix,superscriptaddress]{revtex4-1}
\pdfoutput=1
\usepackage{graphicx}
\usepackage{bm}
\usepackage{subfigure}
\usepackage{url}
\usepackage[hyperindex]{hyperref}
\usepackage{color}
\usepackage[ddmmyy,24hr]{datetime}
\usepackage{bigdelim}
\usepackage{booktabs}
\usepackage{dcolumn}
\usepackage{multirow}
\usepackage{subfigure}
\usepackage{cancel}
\usepackage{stackrel}
\usepackage{paralist}
\usepackage{xspace}
\usepackage{slashed}
\newcommand{\nua}[1]{\ensuremath{\rlap{\kern-2.5pt\ensuremath{\overset{\scriptscriptstyle(-)}{\phantom{\nu}}}}{\ensuremath{{\nu}_{#1}}}}}

\usepackage{enumitem}
\definecolor{darkred}{rgb}{0.5,0,0}
\begin{document}

\title{KATRIN bound on 3+1 active-sterile neutrino mixing and the reactor antineutrino anomaly}

\author{C. Giunti}
\email{carlo.giunti@to.infn.it}
\affiliation{Istituto Nazionale di Fisica Nucleare (INFN), Sezione di Torino, Via P. Giuria 1, I--10125 Torino, Italy}

\author{Y.F. Li}
\email{liyufeng@ihep.ac.cn}
\affiliation{Institute of High Energy Physics,
Chinese Academy of Sciences, Beijing 100049, China}
\affiliation{School of Physical Sciences, University of Chinese Academy of Sciences, Beijing 100049, China}

\author{Y.Y. Zhang}
\email{zhangyiyu@ihep.ac.cn}
\affiliation{Institute of High Energy Physics,
Chinese Academy of Sciences, Beijing 100049, China}
\affiliation{School of Physical Sciences, University of Chinese Academy of Sciences, Beijing 100049, China}

%\date{\dayofweekname{\day}{\month}{\year} \ddmmyydate\today, \currenttime}
\date{15 March 2020}

\begin{abstract}
We present the bounds on 3+1 active-sterile neutrino mixing obtained from the
first results of the KATRIN experiment.
We show that the KATRIN data extend the Mainz and Troitsk bound
to smaller values of $\Delta{m}^2_{41}$
for large mixing
and improves the exclusion of the large-$\Delta{m}^2_{41}$
solution of the Huber-Muller reactor antineutrino anomaly.
We also show that the combined bound
of the Mainz, Troitsk, and KATRIN tritium experiments
and the
Bugey-3,
NEOS,
PROSPECT, and
DANSS
reactor spectral ratio measurements
exclude most of the region in the
($\sin^2\!2\vartheta_{ee},\Delta{m}^2_{41}$)
plane
allowed by the Huber-Muller reactor antineutrino anomaly.
Considering two new calculations of the
reactor neutrino fluxes,
we show that one, that predicts a lower $^{235}\text{U}$ neutrino flux,
is in agreement with the tritium and reactor spectral ratio measurements,
whereas the other leads to a larger tension than the Huber-Muller prediction.
We also show that the combined reactor spectral ratio and tritium measurements
disfavor the Neutrino-4 indication of large active-sterile mixing.
We finally discuss the constraints on the gallium neutrino anomaly.
\end{abstract}

%\pacs{}

\maketitle

\section{Introduction}
\label{sec:introduction}

The KATRIN collaboration presented recently~\cite{Aker:2019uuj} the first results
of their high-precision measurement of the electron spectrum from
$^3\text{H}$ decay near the end point,
where it is sensitive to neutrino masses at the eV level.
They obtained an upper limit of 1.1 eV at 90\% confidence level (CL) for the effective neutrino mass
\begin{equation}
m_{\beta}
=
\sqrt{ \sum_{k=1}^{3} |U_{ek}|^2 m_{k}^2 }
,
\label{mb}
\end{equation}
in the standard three-neutrino mixing framework,
where $U$ is the mixing matrix
and $m_{k}$ is the mass of the neutrino $\nu_{k}$,
with $k=1,2,3$.

The KATRIN collaboration measured the electron spectrum down to
$Q-35\,\text{eV}$,
where $Q \simeq 18.57\,\text{keV}$ is the Q-value of $^3\text{H}$,
that corresponds to the end-point of the electron spectrum in the absence of neutrino mass effects.
Using this spectral measurement, it is possible to constrain also
the mixing with the electron neutrino
of heavier non-standard neutrinos with masses smaller than about 35 eV.
This is interesting in view of the indications in favor of the existence of
such non-standard neutrinos given by the
reactor antineutrino anomaly and the gallium neutrino anomaly
(see the recent reviews in Refs.~\cite{Giunti:2019aiy,Diaz:2019fwt,Boser:2019rta}).
A possible explanation of these anomalies is short-baseline neutrino oscillations
due to the existence of a non-standard neutrino with a mass of the order of 1 eV or larger.
Since it is well established that there are only three active flavor neutrinos,
in the flavor basis the new neutrino must be sterile.
This framework is commonly called 3+1 active-sterile neutrino mixing.

In this paper, we first calculate in Section~\ref{sec:K3}
the upper bound on $m_{\beta}$ in the standard framework of three-neutrino mixing,
in order to test the validity of our analysis of the KATRIN data
by comparing the results with those of the KATRIN collaboration.
Then, in Section~\ref{sec:K3+1},
we calculate the KATRIN bounds on active-sterile neutrino mixing
and we show that they are more stringent than those of the
Mainz~\cite{Kraus:2012he}
and
Troitsk~\cite{Belesev:2012hx,Belesev:2013cba}
experiments discussed in Ref.~\cite{Giunti:2012bc}.
In Section~\ref{sec:RAA},
we compare the KATRIN bounds with the results of the
3+1 analysis of the reactor antineutrino anomaly~\cite{Mention:2011rk}
assuming the standard Huber-Muller reactor neutrino flux prediction~\cite{Mueller:2011nm,Huber:2011wv}
and the two new predictions of
Estienne, Fallot et al.~\cite{Estienne:2019ujo}
and
Hayen, Kostensalo, Severijns, Suhonen~\cite{Hayen:2019eop}.
In Section~\ref{sec:RAA},
we discuss also the bounds of experiments
that measured the reactor antineutrino spectrum at different distances.
In Section~\ref{sec:Nu4},
we compare the positive results of the
Neutrino-4 reactor experiment~\cite{Serebrov:2018vdw}
with the bounds from the tritium experiments and from the other reactor
spectral ratio measurements.
In Section~\ref{sec:gallium}, we discuss the constraints on
the gallium neutrino anomaly.
We finally summarize the results in Section~\ref{sec:conclusions}.

\section{Three neutrino mixing}
\label{sec:K3}

In this section we present the results
of our analysis of the KATRIN data in the standard framework of three-neutrino mixing.
This is useful in order to describe the method that we used in the analysis of the KATRIN data
and in order to check its validity
by comparing the results for $m_{\beta}$ with those obtained by the KATRIN collaboration~\cite{Aker:2019uuj}.

We consider the $\beta$-decay of the gaseous molecular tritium source $ \text{T}_2$:
\begin{equation}\label{key}
\text{T}_2 \rightarrow \ensuremath{\mathrm{^3HeT}}^+ + e^{-} + \bar{\nu}_{e}.
\end{equation}
The differential electron spectrum is given by
\begin{align}
R_\beta(E)
=
&
\frac{G_\mathrm{F}^2 \cos^2\theta_\mathrm{C}}{2 \pi^3}
\,
|\mathcal{M}|^2
\,
F(E ,Z+1)
\nonumber
\\
&
\times
\left( E + m_{e} \right) \sqrt{(E + m_{e})^2 - m^2_{e}}
\nonumber
\\
&
\times
\sum_{i,j}
|U_{ei}|^2
\zeta_j \varepsilon_j \sqrt{\varepsilon_{j}^2 - m^{2}_{i}}
\,
\Theta(\varepsilon_{j}  - m_{i})
,
\label{eq:beta_t2}
\end{align}
where $ G_\mathrm{F} $ is the Fermi constant, $\theta_\mathrm{C}$ is the Cabibbo angle, $\mathcal{M}$ is the nuclear matrix element, %with $ \cos \theta_\mathrm{C} = 0.97425\pm 0.00022 $.
$m_e$ is the electron mass,
$E$ is the kinetic energy of the outgoing electron,
$F(E ,Z+1)$ is the Fermi function describing the Coulomb effect of the electron,
and $Z=1$ is the atomic number of the parent nucleus.
A fully relativistic description of the Fermi function is given by
\begin{equation}
\label{eq:fermi_rel}
F_\text{}(E,Z) = 2(\gamma + 1)
\frac{e^{\pi y}}{(2 p R_{n})^{2(1-\gamma)}}
\frac{\left| \Gamma(\gamma + i y) \right|^2}{ \Gamma(2\gamma\!+\!1)^2},
\end{equation}
where $y = Z\alpha E/p$ and $\gamma = \sqrt{1-\alpha^2 Z^2}$,
with the fine-structure constant $\alpha$ and
the complex Gamma function $\Gamma(z)$~\cite{Konopinski:1935zz}.
The radius of the $ ^3\text{He}^{2+} $ nucleus is
$ R_n = 2.8840\times 10^{-3}/ m_e $~\cite{Ludl:2016ane}.
In Eq.~(\ref{eq:beta_t2}), $\varepsilon_j = E_0-E-V_j$ is the neutrino energy,
with $E_0=M_{\rm T}-M_{\rm {^3He}}-m_{e}$,
where $M_{\rm T}$ and $M_{\rm {^3He}}$ are,
respectively, the mass of the initial and final nucleus.
In the calculation of the $\beta$-decay electron spectrum $R_\beta(E)$, we considered the excitation states of the daughter molecular system, which have
excitation energies $V_j$ and a
final-state distribution with probabilities $\zeta_j$.
These quantities are calculated with the Born-Oppenheimer approximation
and can be found in Refs.~\cite{Saenz:2000dul,Doss:2006zv}.

When the experimental resolution is much larger than the values of neutrino masses, one can define the effective neutrino mass $m_{\beta}$ as in
Eq.~(\ref{mb}) and approximate the differential electron spectrum as
\begin{align}
R_\beta(E)
\simeq
& 
\frac{G_\mathrm{F}^2 \cos^2\theta_\mathrm{C}}{2 \pi^3}
\,
|\mathcal{M}|^2
\,
F(E ,Z+1)
\nonumber
\\
&
\times
\left( E + m_{e} \right) \sqrt{(E + m_{e})^2 - m^2_{e}}
\nonumber
\\
&
\times
\sum_{j} \zeta_j \varepsilon_j \sqrt{\varepsilon_{j}^2 - m^{2}_{\beta}}
\,
\Theta(\varepsilon_{j}  - m_{\beta})
.
\label{eq:beta_t2-app}
\end{align}

The KATRIN experiment combines a windowless gaseous molecular tritium source
%pioneered by the Los Alamos experiment \cite{Robertson:1991vn},
with a spectrometer based on the principle of magnetic adiabatic collimation with electrostatic filtering (MAC-E-filter)~\cite{Lobashev:1985mu,PICARD1992345}.
This apparatus can measure the integral tritium $\beta$-spectrum
\begin{align}
R_\mathrm{} (\langle qU \rangle)
=
&
R_\mathrm{bg} + A_\mathrm{sig} N_\mathrm{T}
\nonumber
\\
&
\times
\int_{qU}^{E_0} R_{\mathrm{\beta}}(E)
\,
f(E - \langle qU \rangle)
\,
d E
 ,
\label{eq:int_spec}
\end{align}
which is the convolution of the differential $\beta$-decay electron spectrum $R_{\mathrm{\beta}}(E)$ with the response function
$f(E - \langle qU \rangle)$.
$N_\mathrm{T}$ denotes the effective number of tritium atoms, $R_\mathrm{bg}$ is the energy-independent background rate and $A_\mathrm{sig}$ is the signal
amplitude.
The response function defines the probability of passing the MAC-E-filter for an electron with the kinetic energy $ E $ at the retarding potential energy $qU$.
$\langle qU \rangle$ is the average over different pixels and scans and serves as the working variable of the integral electron spectrum.
The response function used in our analysis is taken from the red curve of the top panel of Fig.~2 in Ref.~\cite{Aker:2019uuj}.
Note that an energy resolution of $2.8 \ \text{eV}$, which is determined by the energy filter width at the minimal and maximal magnetic fields, has been
included in the response function.
Moreover, an additional Gaussian smearing of $0.25 \ \text{eV}$ is also included to account for the average effect of $\langle qU \rangle$.

For the analysis of the KATRIN data, we considered the $ \chi^2 $ function
\begin{equation}\label{chi2}
\chi^2
=
\sum_{i=1}^{N}
\left(\dfrac{R^\mathrm{obs}_i-R^\mathrm{pred}_i(m_{\beta}^2+\delta{m}_{\beta}^2)}{\sigma_i}\right)^2
+
\left( \dfrac{ \delta{m}_{\beta}^2 }{ 0.32 } \right)^2
,
\end{equation} 
where $R^\mathrm{obs}_i$ and $\sigma_i$ are the experimental rate and its statistical uncertainty
corresponding to each retarding energy value $\langle qU \rangle_i$
in the upper panel of Fig.~3 in Ref.~\cite{Aker:2019uuj}.
$ R^\mathrm{pred}_i$ is the predicted rate calculated according to Eq.~(\ref{eq:int_spec}).
The pull term for the variation $\delta{m}_{\beta}^2$
takes into account the systematic uncertainty of $0.32 \, \text{eV}^2$
on $m_{\beta}^2$
given in Table~I of Ref.~\cite{Aker:2019uuj}.
In the fit we considered four free parameters:
$m^{2}_{\beta}$,
the endpoint $E_{0}$,
the signal amplitude $A_\mathrm{sig}$, and
the background rate $R_\mathrm{bg}$.
We calculated the bounds for $m^{2}_{\beta}$
by marginalizing over
$E_{0}$, $A_\mathrm{sig}$, and $R_\mathrm{bg}$.

In Ref.~\cite{Aker:2019uuj},
the KATRIN collaboration first analyzed the data
allowing negative values of
$m_{\beta}^2$, as discussed in Ref.~\cite{Kleesiek:2018mel}.
With this method,
they obtained
$ m_{\beta}^2 = -1.0 {}^{+0.9}_{-1.1} \ \text{eV}^2 $.
Under the same assumption,
we obtained $ m_{\beta}^2 = -1.0 \pm 0.9 \ \text{eV}^2 $,
which is approximately consistent with the official KATRIN result.

In order to calculate the upper bound on the absolute scale of
neutrino masses in the framework of three-neutrino mixing,
we considered only physical positive values of $m_{\beta}^2$,
as done by the KATRIN collaboration~\cite{Aker:2019uuj}.
We obtained
\begin{equation}
m_{\beta}
<
0.8
\,
(0.9)
\,
\text{eV}
\quad
\text{at}
\quad
90\%
\,
(95\%)
\,
\text{CL}
,
\label{mbeta}
\end{equation}
that nicely coincide with the bounds that the KATRIN collaboration obtained~\cite{Aker:2019uuj}
using the Feldman-Cousins method~\cite{Feldman:1997qc}.

The approximate agreement of our results for $m_{\beta}$
in the standard framework of three-neutrino mixing
with those of the KATRIN collaboration
validates our analysis of the KATRIN data.

\section{3+1 sterile neutrino mixing}
\label{sec:K3+1}

\begin{figure}[!t]
\centering
\includegraphics*[width=\linewidth]{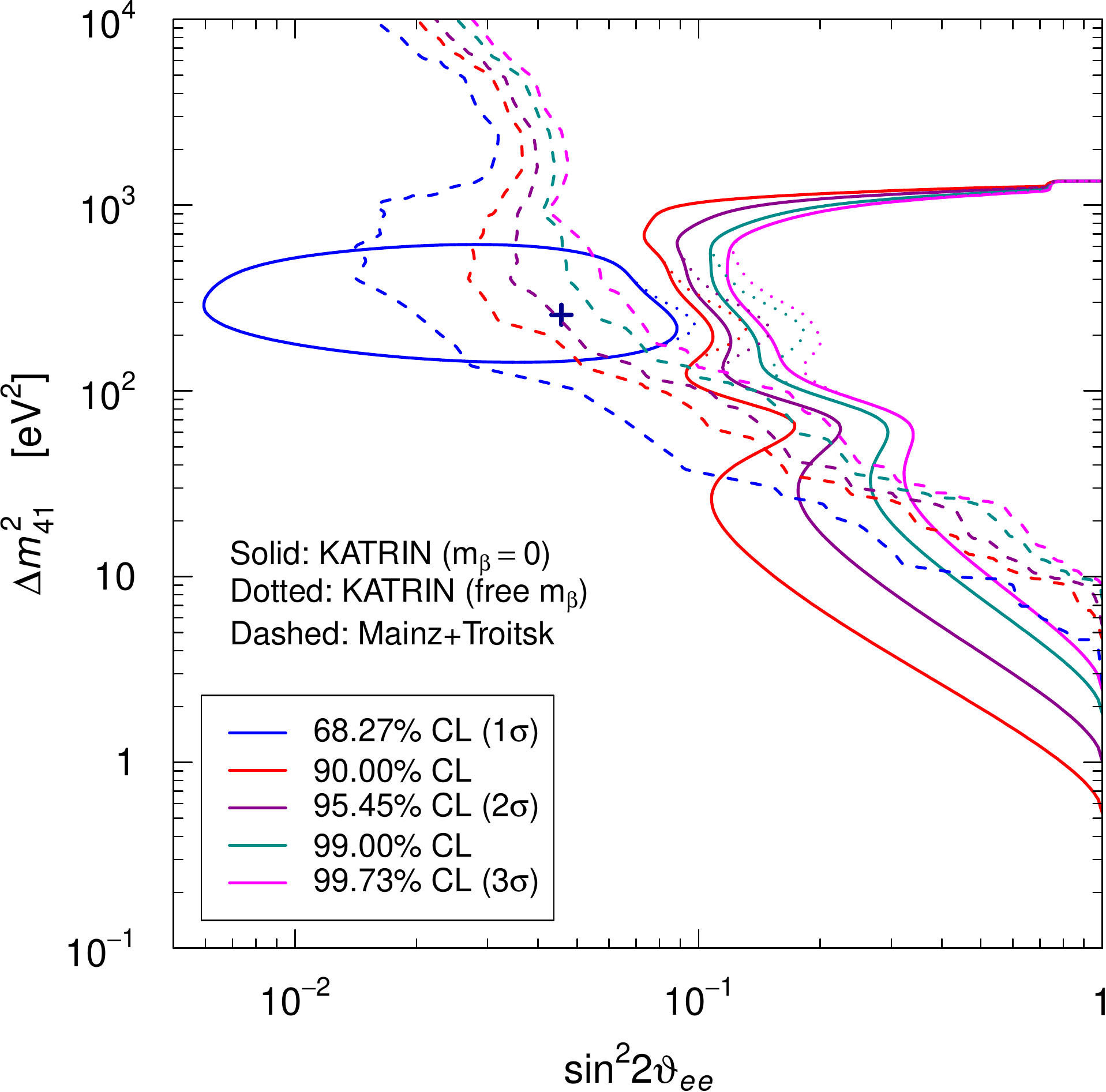}
\caption{ \label{fig:KATRIN}
Confidence level contours in the
($\sin^2\!2\vartheta_{ee},\Delta{m}^2_{41}$)
plane obtained from the analysis of
KATRIN data
with $m_{\beta}=0$ (solid) and free $m_{\beta}$ (dotted),
and from the results of the
Mainz~\cite{Kraus:2012he}
and
Troitsk~\cite{Belesev:2012hx,Belesev:2013cba}
experiments~\cite{Giunti:2012bc}.
The blue cross indicates the KATRIN best-fit point.
}
\end{figure}

After the successful test of our method of analysis of the KATRIN data
in the case of three-neutrino mixing,
we consider the extension to 3+1 active-sterile neutrino mixing with the differential electron spectrum
\begin{equation}
R_\beta(E)
=
( 1 - |U_{e4}|^2 ) \, R_\beta(E,m_{\beta})
+
|U_{e4}|^2 \, R_\beta(E,m_{4})
,
\label{Rbeta3+1}
\end{equation}
where
$U$ is the $4\times4$ unitary mixing matrix,
$R_\beta(E,m_{\beta})$ is the three-neutrino
differential electron spectrum in Eq.~(\ref{eq:beta_t2-app})
with $m_{\beta}$ redefined by\footnote{
The necessity of the factor $( 1 - |U_{e4}|^2 )$
in front of $R_\beta(E,m_{\beta})$
can be understood by noting that in the limit of negligible masses of
$\nu_{1}$,
$\nu_{2}$, and
$\nu_{3}$,
their contribution to the electron spectrum
is given by Eq.~(\ref{eq:beta_t2}) with
$m_{1}=m_{2}=m_{3}=0$.
In this case one can extract a common mixing factor
$ \sum_{i=1}^{3} |U_{ei}|^2 = 1 - |U_{e4}|^2 $
and write the contribution of
$\nu_{1}$,
$\nu_{2}$, and
$\nu_{3}$
as
$( 1 - |U_{e4}|^2 ) \, R_\beta(E,0) $.
}
\begin{equation}
m_{\beta}^2
=
\sum_{k=1}^{3} \dfrac{|U_{ek}|^2}{1-|U_{e4}|^2} \, m_{k}^2
,
\label{mb1}
\end{equation}
and
$R_\beta(E,m_{4})$ has the same expression with $m_{\beta}$ replaced by $m_{4}$.
We will compare the results of our analysis of the KATRIN data
with the results of short-baseline (SBL) reactor neutrino oscillation experiments,
that probe the effective SBL survival probability
\begin{equation}
P^{\text{SBL}}_{\nua{e}\to\nua{e}}
=
1 - \sin^2\!2\vartheta_{ee} \, \sin^2\!\left( \dfrac{\Delta{m}^2_{41} L}{4 E} \right)
,
\label{pee}
\end{equation}
where
$\Delta{m}^2_{ij}=m_{i}^2-m_{j}^2$,
$\sin^2\!2\vartheta_{ee} = 4 |U_{e4}|^2 ( 1 - |U_{e4}|^2 )$,
$L$ is the source-detector distance, and $E$ is the neutrino energy.
Note that neutrino oscillation experiments are sensitive to the squared-mass difference\footnote{
The effective SBL survival probability (\ref{pee})
is derived under the approximation
$\Delta{m}^2_{41} \simeq \Delta{m}^2_{42} \simeq \Delta{m}^2_{43} \gtrsim 0.1 \, \text{eV}^2$,
taking into account the smallness of the values of $\Delta{m}^2_{21}$
and
$|\Delta{m}^2_{31}| \simeq |\Delta{m}^2_{32}|$,
given in Eqs.~(\ref{d21}) and (\ref{d31}), respectively.
}
$\Delta{m}^2_{41} \simeq \Delta{m}^2_{42} \simeq \Delta{m}^2_{43}$,
whereas the KATRIN experiment is sensitive to
$m_{\beta}$ and $m_{4}$.
Therefore, in order to compare the respective results one must make some assumption
on the value of one of the three light neutrino masses ($m_{1}$, $m_{2}$, $m_{3}$),
that fixes the value of $m_{\beta}$ through the precise knowledge of the values of the
three-neutrino mixing parameters
obtained by global fits of solar, atmospheric and long-baseline neutrino oscillation data~\cite{deSalas:2017kay,Capozzi:2018ubv,Esteban:2018azc,Tanabashi:2018oca}:
\begin{align}
\null & \null
\Delta{m}^2_{21} \simeq 7.5 \times 10^{-5} \, \text{eV}^2
,
\label{d21}
\\
\null & \null
|\Delta{m}^2_{31}| \simeq |\Delta{m}^2_{32}| \simeq 2.5 \times 10^{-3} \, \text{eV}^2
,
\label{d31}
\\
\null & \null
|U_{e2}|^2 \simeq 0.3
,
\label{ue2}
\\
\null & \null
|U_{e3}|^2 \simeq 0.022
,
\label{ue3}
\end{align}
with positive and negative $ \Delta{m}^2_{31} \simeq \Delta{m}^2_{32} $
in the two possible cases of Normal Ordering (NO) and Inverted Ordering (IO)
of the three light neutrino masses, respectively
(see the recent review in Ref.~\cite{deSalas:2018bym}).
Equation~(\ref{mb1}) can be written as
\begin{equation}
m_{\beta}^2
=
m_{1}^2
+
\dfrac{|U_{e2}|^2 \Delta{m}^2_{21} + |U_{e3}|^2 \Delta{m}^2_{31}}{1-|U_{e4}|^2}
.
\label{mb2}
\end{equation}
Hence, we have
\begin{align}
\text{NO:}
\quad
m_{\beta}^2
\null & \null
\simeq
m_{1}^2
+
\dfrac{7.8 \times 10^{-5} \, \text{eV}^2}{1-|U_{e4}|^2}
,
\label{NO}
\\
\text{IO:}
\quad
m_{\beta}^2
\null & \null
\simeq
m_{1}^2
-
\dfrac{3.3 \times 10^{-5} \, \text{eV}^2}{1-|U_{e4}|^2}
.
\label{IO}
\end{align}
Therefore, taking into account that the sensitivity of KATRIN
to $m_{\beta}^2$ is at the level of the $\text{eV}^2$
and considering\footnote{
Since 3+1 active-sterile neutrino mixing is allowed only as a perturbation
of standard three-neutrino mixing,
$|U_{e4}|^2$ cannot be large
(see the recent reviews in Refs.~\cite{Giunti:2019aiy,Diaz:2019fwt,Boser:2019rta}).
}
$|U_{e4}|^2<0.5$ for $\sin^2\!2\vartheta_{ee}<1$,
we can neglect the small deviations of $m_{\beta}^2$ from $m_{1}^2$
in Eqs.~(\ref{NO}) and (\ref{IO}),
and consider the approximate relation
\begin{equation}
\Delta{m}^2_{41}
\simeq
m_{4}^2 - m_{\beta}^2
.
\label{d41mb}
\end{equation}
We performed two analyses of the KATRIN data in the framework of
3+1 active-sterile neutrino mixing.
First,
we fitted the data considering
$A_\mathrm{sig}$, $R_\mathrm{bg}$, $E_{0}$, $m_{\beta}$, $|U_{e4}|^2$, and $m_{4}$ as free parameters
and we calculated the ``free $m_{\beta}$'' confidence level contours in the
($\sin^2\!2\vartheta_{ee},\Delta{m}^2_{41}$)
plane shown in Figure~\ref{fig:KATRIN}
marginalizing the $\chi^2$ over $A_\mathrm{sig}$, $R_\mathrm{bg}$, $E_{0}$ and $m_{\beta}$.
This is the most general bound on 3+1 mixing given by the KATRIN data.
We also calculated the confidence level contours in the case of a negligible $m_{\beta}$,
shown by the $m_{\beta}=0$ lines in Figure~\ref{fig:KATRIN}.
This is a reasonable assumption motivated by the likeliness of a neutrino mass hierarchy,
with $m_{1,2,3} \ll m_{4}$.
It is also useful for the comparison in Figure~\ref{fig:KATRIN}
of the KATRIN bounds with the exclusion curves of the
Mainz~\cite{Kraus:2012he}
and
Troitsk~\cite{Belesev:2012hx,Belesev:2013cba}
experiments obtained in Ref.~\cite{Giunti:2012bc} under the same assumption.
One can see from Figure~\ref{fig:KATRIN} that the KATRIN bounds obtained with free $m_{\beta}$
and $m_{\beta}=0$ are slightly different only around
$\Delta{m}^2_{41} \approx 200-300 \, \text{eV}^2$,
where the Mainz+Troitsk bound is dominant.
Therefore,
in the following we can safely consider only the analysis of KATRIN data with $m_{\beta}=0$.

\begin{figure}[!t]
\centering
\includegraphics*[width=\linewidth]{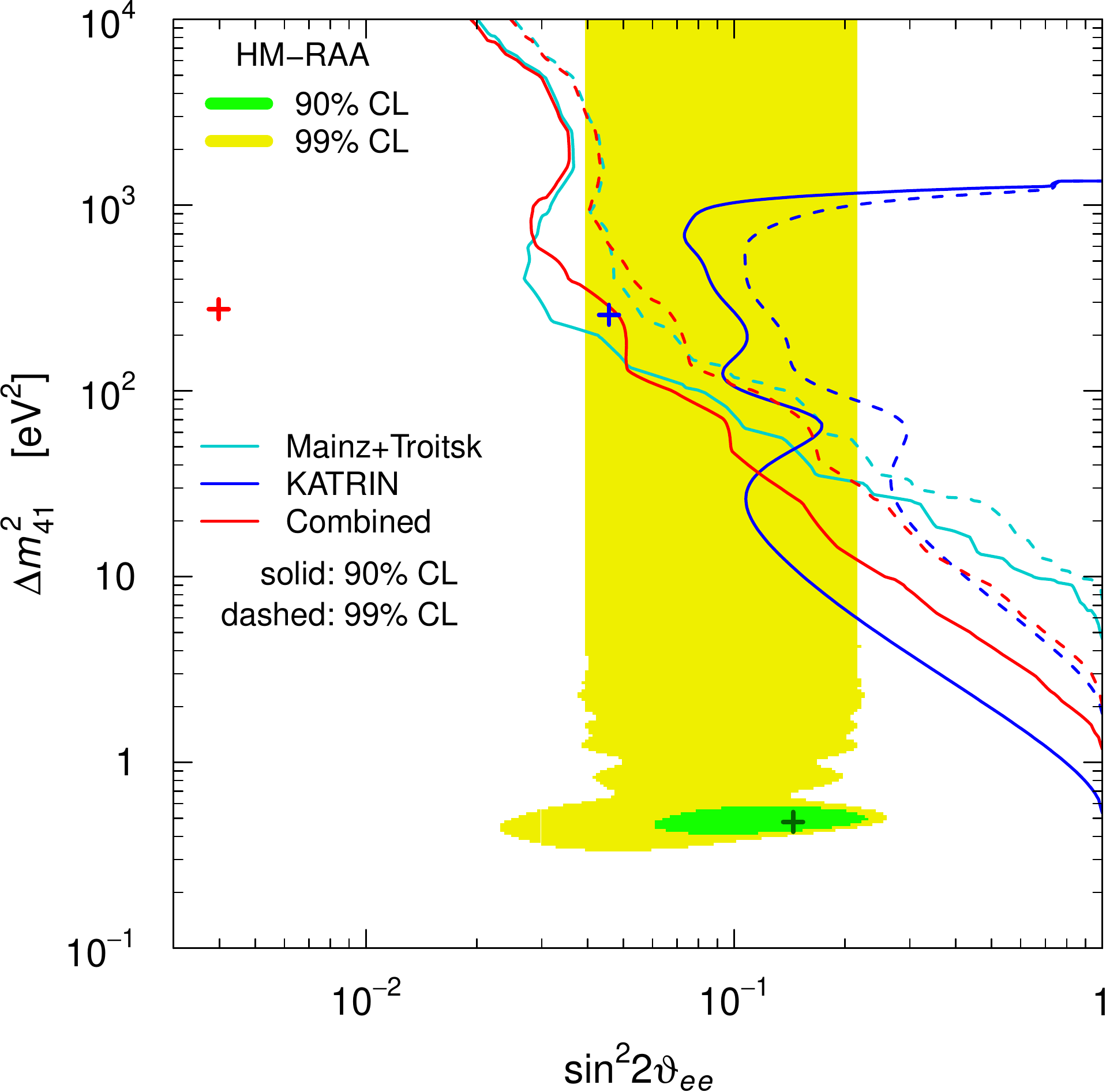}
\caption{ \label{fig:KMT}
90\% and 99\%~CL exclusion curves in the
($\sin^2\!2\vartheta_{ee},\Delta{m}^2_{41}$)
plane obtained from the analysis of
KATRIN data
with free $m_{\beta}$ and $m_{\beta}=0$.
Also shown are the exclusion curves of the
Mainz~\cite{Kraus:2012he}
and
Troitsk~\cite{Belesev:2012hx,Belesev:2013cba}
experiments obtained in Ref.~\cite{Giunti:2012bc}
and the combined exclusion curves.
The green and yellow regions are allowed at
90\% and 99\%~CL by the neutrino oscillation solution~\cite{Giunti:2019qlt}
of the Huber-Muller reactor antineutrino anomaly (HM-RAA).
The crosses indicate the best-fit points.
}
\end{figure}

Figure~\ref{fig:KATRIN} shows that the analysis of the KATRIN data yields a
$1\sigma$ allowed region with non-zero active-sterile neutrino mixing
around the best-fit point at
$
\sin^2\!2\vartheta_{ee}
=
0.046
$
and
$
\Delta{m}^2_{41}
=
257 \, \text{eV}^2
$.
Since the $1\sigma$ KATRIN allowed region is in tension with the Mainz+Troitsk bound,
it is likely due to a statistical fluctuation.
Therefore,
in the following we will consider only
the KATRIN exclusion curves at higher confidence level.
However,
the slight tension between KATRIN and Mainz+Troitsk
has effects on the combined bound that will be discussed below.

Figure~\ref{fig:KATRIN} shows that the KATRIN data allow us to extend the
Mainz+Troitsk excluded region at large mixing to smaller value of
$\Delta{m}^2_{41}$,
reaching the interesting values of $\Delta{m}^2_{41}$ below $10 \, \text{eV}^2$.
In the logarithmic scale of Figure~\ref{fig:KATRIN},
the KATRIN bounds on $\sin^2\!2\vartheta_{ee}$ have an approximately linear decrease
when $\Delta{m}^2_{41}$ increases from about $6 \, \text{eV}^2$
to about $30 \, \text{eV}^2$,
that corresponds to $ m_{4} \approx 5.5 \, \text{eV} $.
For larger values of $\Delta{m}^2_{41}$,
the effect of $\nu_{4}$ on the electron spectrum
occurs at a distance from the end point for which the data are
less constraining.
This leads to oscillations of the bounds from
$\Delta{m}^2_{41} \approx 30 \, \text{eV}^2$
to
$\Delta{m}^2_{41} \approx 10^{3} \, \text{eV}^2$,
that corresponds to the value $m_{4} \approx 32 \, \text{eV}$
for which the data become completely ineffective.

Figure~\ref{fig:KMT}
shows the combined 90\% and 99\% CL bounds of the tritium experiments
compared with the corresponding KATRIN and Mainz+Troitsk bounds
(and the regions allowed by the reactor antineutrino anomaly
to be discussed in Section~\ref{sec:RAA}).
One can see that the combined tritium bound extends the
Mainz+Troitsk excluded region at large mixing to values of
$\Delta{m}^2_{41}$ below $10 \, \text{eV}^2$.
However, the combined tritium bound is less stringent than the
KATRIN bound in the small-$\Delta{m}^2_{41}$ range where the KATRIN data are dominant
(this occurs much more for the 90\% CL bound than for the 99\% CL bound).
This strange behavior is due to the location of the minimum of the
KATRIN $\chi^2$, that is shown in Figure~\ref{fig:KMT} by the blue cross,
in a point where the Mainz+Troitsk $\chi^2$ is not small.
The location of the minimum $\chi^2$ of the combined fit
is shown in Figure~\ref{fig:KMT} by the red cross and it is obviously larger
than the KATRIN $\chi^2$ minimum
($\Delta\chi^2=2.5$),
since it lies out of the KATRIN $1\sigma$ allowed region shown in Fig.~\ref{fig:KATRIN}.
Since the confidence level contours are determined by the difference of
$\chi^2$
(given by Table~39.2 of Ref.~\cite{Tanabashi:2018oca} for two degrees of freedom)
with respect to the minimum corresponding to the chosen confidence level,
the increase of the $\chi^2$ minimum
leads to a shift towards larger values of
$\sin^2\!2\vartheta_{ee}$
of the combined tritium bound with respect to the KATRIN bound
in the $\Delta{m}^2_{41}$ range where KATRIN is dominant.
For a similar reason,
the combined tritium bound is less stringent than the Mainz+Troitsk bound for
$\Delta{m}^2_{41} \approx 100-600 \, \text{eV}^2$,
where the Mainz+Troitsk bound is dominant.

\begin{figure}[!t]
\centering
\includegraphics*[width=\linewidth]{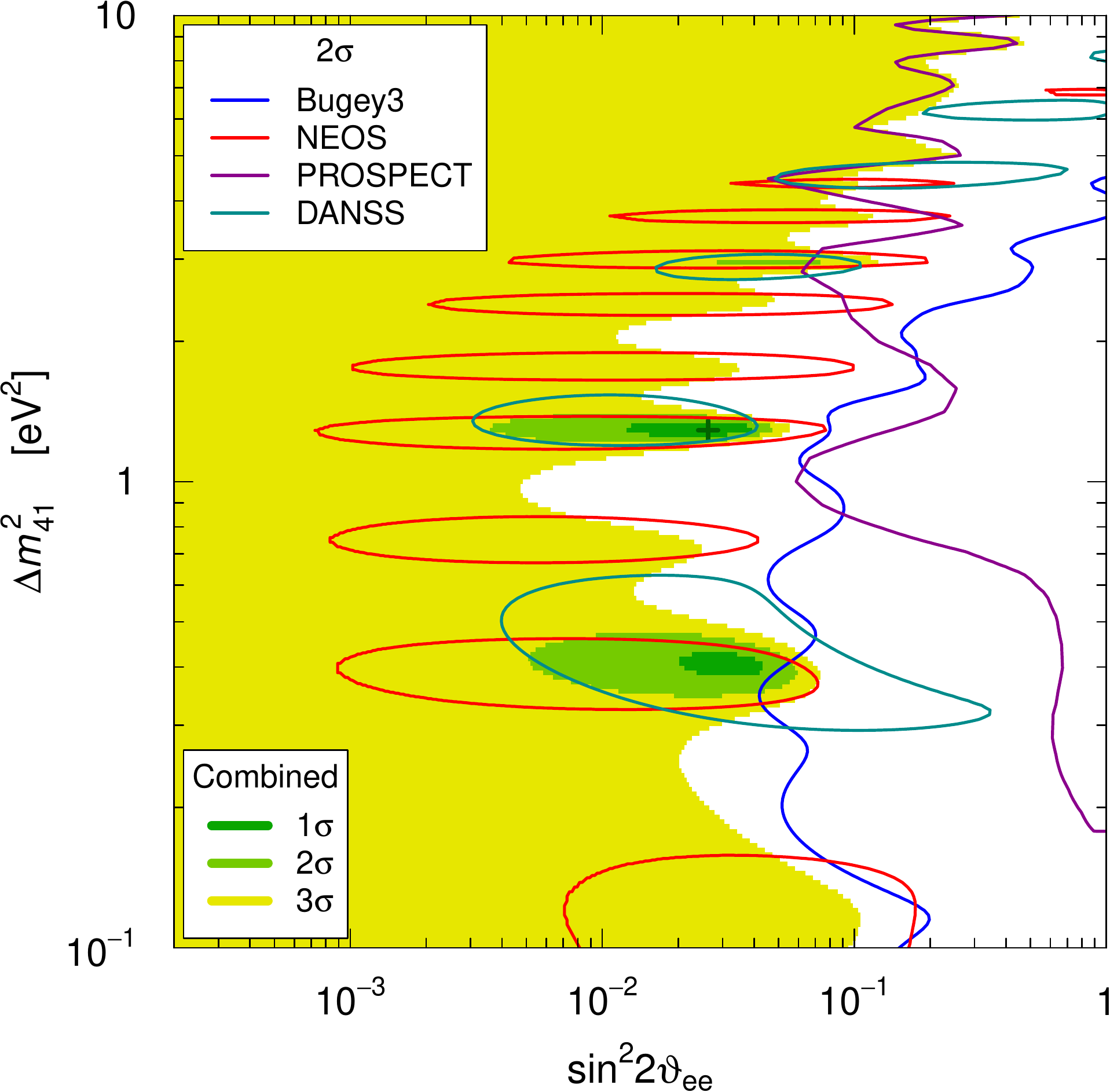}
\caption{ \label{fig:RSR}
Contours of the $2\sigma$ regions in the
($\sin^2\!2\vartheta_{ee},\Delta{m}^2_{41}$)
plane obtained from the reactor spectral ratio measurements of the
Bugey-3, NEOS, PROSPECT and DANSS experiments.
The shadowed regions are allowed by the combined fit,
with the best fit point indicated by the cross.
}
\end{figure}

\begin{figure}[!t]
\centering
\includegraphics*[width=\linewidth]{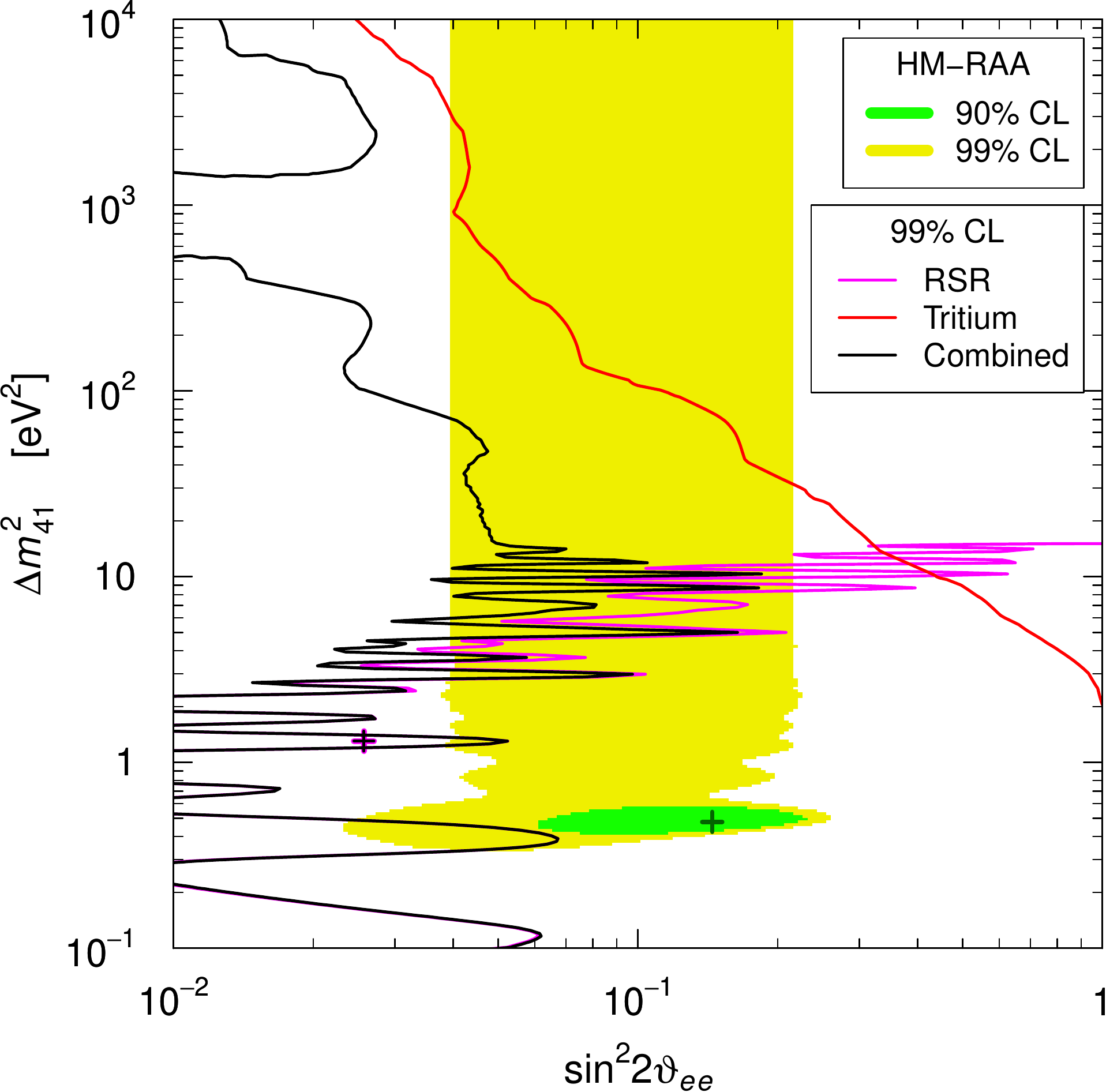}
\caption{ \label{fig:T+R}
99\%~CL exclusion curves in the
($\sin^2\!2\vartheta_{ee},\Delta{m}^2_{41}$)
plane obtained from the analysis of the data of
the Mainz, Troitsk and KATRIN tritium experiments
and the combined analysis of the reactor spectral ratio (RSR) measurements of the
Bugey-3, NEOS, PROSPECT and DANSS experiments.
Also shown is the combined tritium and reactor spectral-ratio exclusion curve
and the regions allowed at 90\% and 99\%~CL by the neutrino oscillation solution~\cite{Giunti:2019qlt}
of the Huber-Muller reactor antineutrino anomaly (HM-RAA).
The crosses indicate the best-fit points.
}
\end{figure}

\section{The reactor antineutrino anomaly}
\label{sec:RAA}

In Figure~\ref{fig:KMT} we have also drawn the regions allowed by the reactor antineutrino anomaly (HM-RAA)~\cite{Mention:2011rk}
according to the recent analysis in Ref.~\cite{Giunti:2019qlt} of reactor antineutrino data
compared with the Huber-Muller prediction~\cite{Mueller:2011nm,Huber:2011wv}
(see also Ref.~\cite{Berryman:2019hme}).
One can see that the combined constraints of tritium-decay experiments
can exclude the large-$\Delta{m}^2_{41}$ part of the RAA 99\% allowed region,
but it is still too weak to affect the 90\% allowed region around the best-fit point.
Note that this HM-RAA region is different from the original reactor antineutrino anomaly
allowed region in Ref.~\cite{Mention:2011rk} (see also Ref.~\cite{Giunti:2012tn})
mainly because it takes into account only the measured reactor neutrino rates,
without the Bugey-3~\cite{Declais:1995su} 14 m / 15 m spectral ratio that were included in
Refs.~\cite{Mention:2011rk,Giunti:2012tn}.
As nicely illustrated in Fig.~1 of Ref.~\cite{Kopp:2013vaa},
the Bugey-3 spectral ratio excludes large mixing for
$\Delta{m}^2_{41} \lesssim 2 \, \text{eV}^2$,
moving the best-fit region from
$\Delta{m}^2_{41} \approx 0.5 \, \text{eV}^2$
to
$\Delta{m}^2_{41} \approx 1.8 \, \text{eV}^2$.
However, in discussing the reactor antineutrino anomaly it is better to separate the
model-dependent anomaly based on the absolute neutrino rate measurements and
the model-independent implications of the spectral-ratio measurements.

Recently,
also the new reactor neutrino experiments
DANSS~\cite{Alekseev:2018efk,Danilov:2019aef},
PROSPECT~\cite{Ashenfelter:2018iov}, and
STEREO~\cite{Almazan:2018wln,AlmazanMolina:2019qul}
measured the reactor antineutrino spectrum at different distances.
Moreover, the NEOS~\cite{Ko:2016owz} experiments presented the results of a measurement of
the reactor antineutrino spectrum at 24 m from a reactor,
relative to the spectrum measured at about 500 m by the Daya Bay near detectors~\cite{An:2016srz}.
These measurements provide information on short-baseline neutrino oscillations that
are independent of the theoretical calculation of the reactor antineutrino flux.
Therefore, they can test the model-dependent reactor antineutrino anomaly
and their results can be combined with the bounds given by the tritium experiments.
Here we consider the published results of the
Bugey-3~\cite{Declais:1995su},
NEOS~\cite{Ko:2016owz}, and
PROSPECT~\cite{Ashenfelter:2018iov} experiments,
together with the preliminary 2019 results of the DANSS~\cite{Danilov:2019aef} experiment,
that improve significantly the published 2018 results~\cite{Danilov:2019aef}.
We cannot include in the analysis the results of the STEREO~\cite{Almazan:2018wln,AlmazanMolina:2019qul}
experiment, because there is not enough available information.
For the Bugey-3 experiment we used the same analysis
that we used in previous papers~\cite{Giunti:2019qlt,Gariazzo:2017fdh,Gariazzo:2018mwd}.
For the NEOS experiment we use the $\chi^2$ table
kindly provided by the NEOS collaboration.
For the PROSPECT experiment we use the $\chi^2$ table
published as ``Supplemental Material'' of Ref.~\cite{Ashenfelter:2018iov}.
For the DANSS experiment we performed an approximate least-square analysis of the 2019
data presented in Fig.~5 of Ref.~\cite{Danilov:2019aef} that reproduces approximately the
DANSS exclusion curves in Fig.~6 of the same paper.

Figure~\ref{fig:RSR} shows the contours of the $2\sigma$ regions in the
($\sin^2\!2\vartheta_{ee},\Delta{m}^2_{41}$)
plane obtained from the reactor spectral ratio measurements of the
Bugey-3, NEOS, PROSPECT and DANSS experiments,
and the regions allowed at $1\sigma$, $2\sigma$, and $3\sigma$
by the combined fit.
One can see that there is an indication in favor of
short-baseline oscillations at the level of about $2\sigma$,
that is due to the coincidence of the NEOS and DANSS allowed regions at
$
\Delta{m}^2_{41}
\approx
1.3
\, \text{eV}^2
$,
where there is the best-fit point of the combined fit for
$
\sin^2\!2\vartheta_{ee}
=
0.026
$
and
$
\Delta{m}^2_{41}
=
1.3
\, \text{eV}^2
$.
The NEOS and DANSS allowed regions partially overlap also at
$
\Delta{m}^2_{41}
\approx
0.4
\, \text{eV}^2
$,
where there is a combined $1\sigma$-allowed region, and at
$
\Delta{m}^2_{41}
\approx
3
\, \text{eV}^2
$,
where there is a tiny combined $2\sigma$-allowed region.
This model-independent indication in favor of short-baseline oscillations
was discussed in Refs.~\cite{Gariazzo:2018mwd,Dentler:2018sju}
using the 2018~\cite{Alekseev:2018efk} DANSS data
and in Ref.~\cite{Berryman:2019hme}
using both the 2018 and the 2019~\cite{Danilov:2019aef} DANSS data.
Here, as explained above, we use the 2019 DANSS data,
that lead to a diminished indication
in favor of short-baseline oscillations
with respect to the 2018 DANSS data.
Indeed, from the combined NEOS and DANSS analyses we find only a
$2.6\sigma$
indication of short-baseline oscillations,
that is smaller than the $3.7\sigma$ obtained in Ref.~\cite{Gariazzo:2018mwd}.
These values agree approximately with those found in Ref.~\cite{Berryman:2019hme}.

Figure~\ref{fig:T+R} shows the 99\% exclusion curve in the
($\sin^2\!2\vartheta_{ee},\Delta{m}^2_{41}$)
plane obtained from the combined analysis of the
Bugey-3, NEOS, PROSPECT and DANSS spectral ratios,
that constrain the mixing for low values of $\Delta{m}^2_{41}$,
together with the combined 99\%~CL exclusion curve of the
Mainz, Troitsk and KATRIN tritium experiments,
that constrains the mixing for large values of $\Delta{m}^2_{41}$.
Figure~\ref{fig:T+R} shows also the combined
tritium and reactor spectral-ratio 99\%~CL exclusion curve,
that disfavors most of the 99\%~CL allowed region~\cite{Giunti:2019qlt}
of the Huber-Muller reactor antineutrino anomaly.
Note that the combined tritium and reactor spectral-ratio bound at large values of
$\Delta{m}^2_{41}$
is much more stringent than the tritium bound,
in spite of the lack of sensitivity of the reactor spectral ratio experiments
for $\Delta{m}^2_{41} \gtrsim 10 \, \text{eV}^2$.
The reason is that the
99\% exclusion curve is determined by the appropriate difference of
$\chi^2$
(given by Table~39.2 of Ref.~\cite{Tanabashi:2018oca} for two degrees of freedom)
with respect to the global $\chi^2$ minimum that occurs at
$
\sin^2\!2\vartheta_{ee}
=
0.026
$
and
$
\Delta{m}^2_{41}
=
1.3
\, \text{eV}^2
$.
This point practically coincides with the reactor spectral ratio best fit in Figure~\ref{fig:RSR} and is far away from the tritium-only best fit in Figure~\ref{fig:KMT},
because the reactor spectral ratio data are dominant in the combined fit.
The stringent combined tritium and reactor spectral-ratio bound at large values of
$\Delta{m}^2_{41}$ is due to the combined effects of the tritium bound in that region and the reactor spectral ratio data that allow these large values of $\Delta{m}^2_{41}$ only at $3\sigma$,
as can be seen in Figure~\ref{fig:RSR}.
The reactor spectral ratio data $\chi^2$ has this effect for large values of $\Delta{m}^2_{41}$,
where the experiments are not sensitive,
because the data prefer the small-$\Delta{m}^2_{41}$ region near the best fit.
This is an obviously correct effect if one thinks that the measurement of a physical quantity
in an experiment
excludes all the values of the physical quantity that are enormously different from the measured one
and for which the experiment is not sensitive.

\begin{figure}[!t]
\centering
\includegraphics*[width=\linewidth]{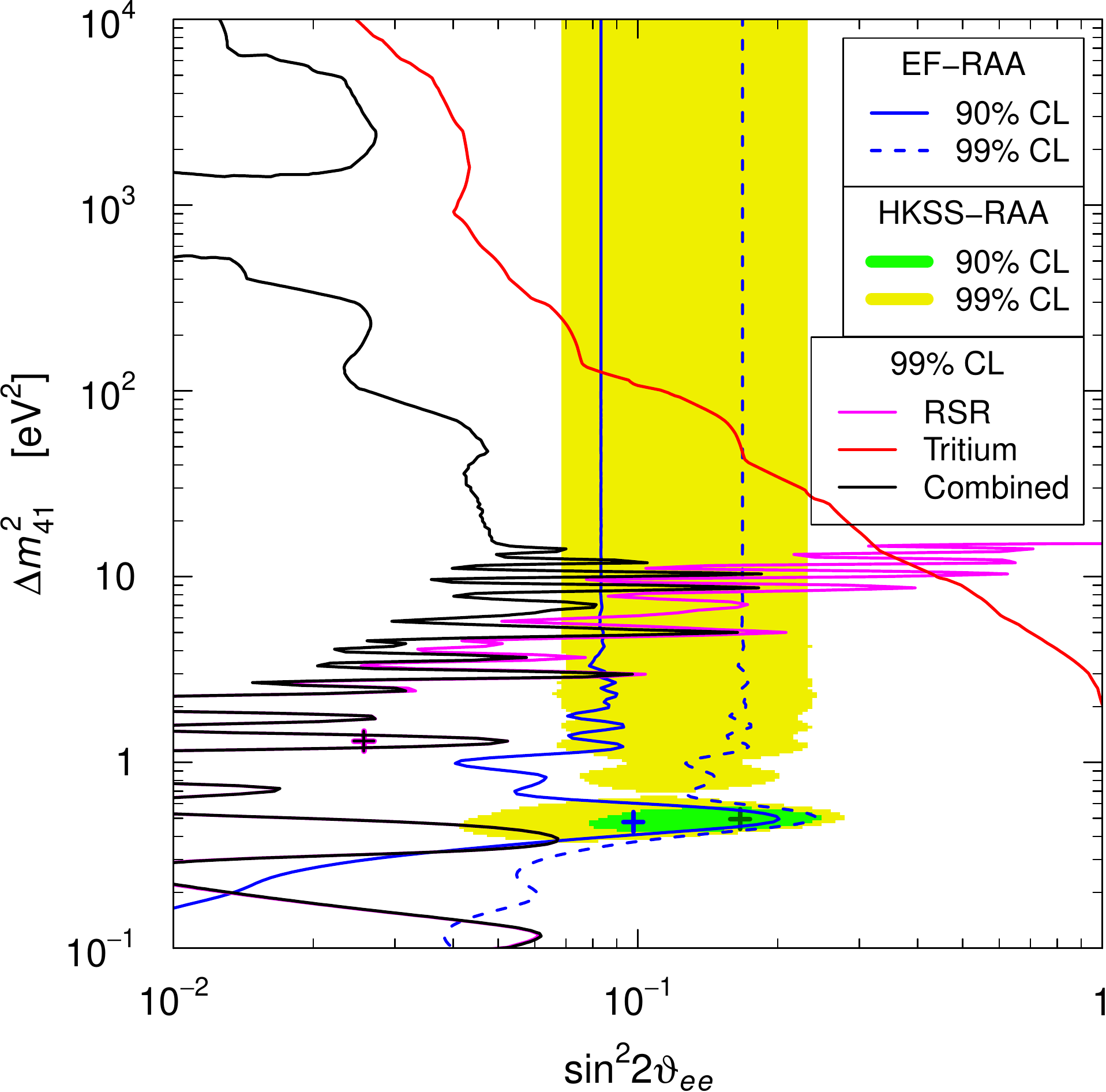}
\caption{ \label{fig:EF-HKSS}
99\%~CL exclusion curves in the
($\sin^2\!2\vartheta_{ee},\Delta{m}^2_{41}$)
plane obtained from the combined  analysis of the data of
the Mainz, Troitsk and KATRIN tritium experiments
and the combined analysis of the reactor spectral ratio (RSR) measurements of the
Bugey-3, NEOS, PROSPECT and DANSS experiments.
Also shown is the combined tritium and reactor spectral-ratio exclusion curve
and the regions allowed at 90\% and 99\%~CL by the fits of the
absolute reactor rates assuming
the Estienne, Fallot et al.~\cite{Estienne:2019ujo} (EF-RAA)
and
the Hayen, Kostensalo, Severijns, Suhonen~\cite{Hayen:2019eop} (HKSS-RAA)
reactor neutrino fluxes.
The crosses indicate the best-fit points.
}
\end{figure}

Figure~\ref{fig:T+R} shows that
there is a tension between the active-sterile oscillations
indicated by the Huber-Muller reactor antineutrino anomaly
and the combined bound obtained from tritium and reactor spectral-ratio measurements.
However,
it is likely that the Huber-Muller neutrino flux prediction must be revised,
as indicated by the observation of a large spectral distortion at 5 MeV in the
RENO \cite{Seo:2014xei,RENO:2015ksa},
Double Chooz \cite{Abe:2014bwa},
Daya Bay \cite{An:2016srz},
and
NEOS \cite{Ko:2016owz}
experiments
(see the reviews in Refs.~\cite{Huber:2016fkt,Hayes:2016qnu}).
As already discussed in Ref.~\cite{Berryman:2019hme},
there are two recent reactor neutrino flux calculations
that may improve the Huber-Muller prediction:
the calculation of Estienne, Fallot et al. (EF)~\cite{Estienne:2019ujo}
that is based on the summation method,
and the calculation of Hayen, Kostensalo, Severijns, Suhonen (HKSS)~\cite{Hayen:2019eop}
that improves the conversion method by including the effects
of forbidden $\beta$ decays through shell-model calculations.
Unfortunately,
as discussed in Ref.~\cite{Berryman:2019hme},
a comparison of the results of the two new calculations does not lead to a clarification of the problem
of the reactor antineutrino anomaly,
because the corresponding neutrino flux predictions diverge:
the EF calculation resulted in a $^{235}\text{U}$
neutrino flux prediction that is smaller than the HM prediction,
leading to a decrease of the reactor antineutrino anomaly,
whereas the HKSS fluxes are larger than the HM fluxes,
leading to an increase of the reactor antineutrino anomaly.
Figure~\ref{fig:EF-HKSS}
show a comparison of the bounds in the
($\sin^2\!2\vartheta_{ee},\Delta{m}^2_{41}$)
plane obtained from the tritium experiments
and the reactor spectral ratios
with the regions allowed by the fits of the absolute reactor rates
assuming the EF and HKSS fluxes.
We took into account the uncertainties of the HKSS fluxes given in Ref.~\cite{Hayen:2019eop}.
On the other hand, since the EF cross section per fission are given in Ref.~\cite{Estienne:2019ujo}
without the associated uncertainties, for them we adopted the
uncertainties associated with the summation spectra estimated in Ref.~\cite{Hayes:2017res}:
5\% for
$^{235}\text{U}$,
$^{239}\text{Pu}$, and
$^{241}\text{Pu}$,
and 10\% for $^{238}\text{U}$.

From Figure~\ref{fig:EF-HKSS}, one can see that the EF neutrino flux calculation
leads only to an upper bound on the mixing at 90\%~CL and higher.
Therefore, in this case the reactor antineutrino anomaly is not statistically significant
and the EF-RAA upper bound is compatible with the upper bounds obtained
from the tritium experiments
and the reactor spectral ratios.

On the other hand,
the HKSS fluxes lead to an increase of the reactor antineutrino anomaly
with respect to the HM prediction and the corresponding HKSS-RAA allowed regions in
Figure~\ref{fig:EF-HKSS}
are limited to larger mixing than the HM-RAA allowed regions in Figure~\ref{fig:T+R}.
Therefore, the tension of the HKSS-RAA with the
tritium and reactor spectral ratios bounds is larger than that of the HM-RAA.
From Figure~\ref{fig:EF-HKSS} one can see that only very small portions of the
HKSS-RAA 99\% allowed region
are not excluded by the combined 99\% bound of the tritium experiments
and the reactor spectral ratios.

\begin{figure}[!t]
\centering
\includegraphics*[width=\linewidth]{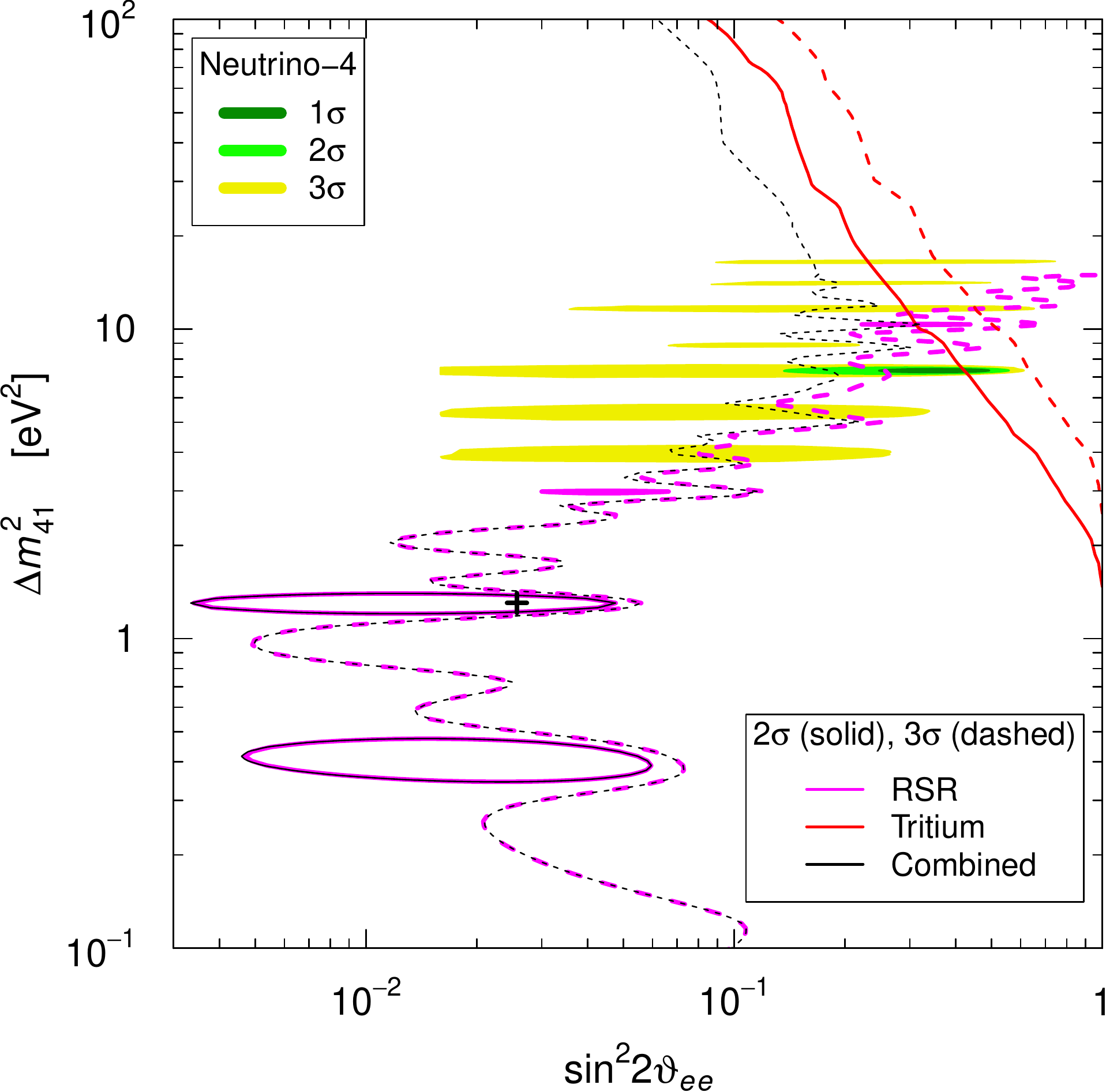}
\caption{ \label{fig:nu4}
$2\sigma$ and $3\sigma$ bounds in the
($\sin^2\!2\vartheta_{ee},\Delta{m}^2_{41}$)
plane obtained from the combined analysis of the data of
the Mainz, Troitsk and KATRIN tritium experiments
and the combined analysis of the reactor spectral ratio (RSR) measurements of the
Bugey-3, NEOS, PROSPECT and DANSS experiments.
Also shown are the combined tritium and reactor spectral-ratio bounds
(with the best fit indicated by the black cross)
and the regions allowed at $1\sigma$, $2\sigma$, and $3\sigma$
by the results of the Neutrino-4 reactor experiment~\cite{Serebrov:2018vdw}.
}
\end{figure}

\section{Neutrino-4}
\label{sec:Nu4}

Let us now consider the results of the Neutrino-4 reactor experiment~\cite{Serebrov:2018vdw},
that is another experiment that measured the ratios of the spectra at different distances from the reactor,
between 6 and 12 m.
We did not consider it so far because the result of this experiment
is an anomalous indication of short-baseline oscillations
with large mixing that is in tension with all the other experimental results.
This can be seen in Figure~\ref{fig:nu4},
where
we compare the bounds in the 
($\sin^2\!2\vartheta_{ee},\Delta{m}^2_{41}$)
plane obtained from the tritium experiments
and the reactor spectral ratios
with the allowed regions of the Neutrino-4 reactor experiment~\cite{Serebrov:2018vdw}.
One can see that
the large-mixing parts of the Neutrino-4 allowed regions are excluded
by the $3\sigma$ combined tritium and reactor spectral-ratio exclusion curve.

At $2\sigma$,
the combination of the reactor spectral-ratio and tritium measurements
have allowed regions at
$
\Delta{m}^2_{41}
\approx
1.3
\, \text{eV}^2
$,
where there is the best-fit point for
$
\sin^2\!2\vartheta_{ee}
=
0.026
$
and
$
\Delta{m}^2_{41}
=
1.3
\, \text{eV}^2
$,
and at
$
\Delta{m}^2_{41}
\approx
0.4
\, \text{eV}^2
$,
that correspond to those in Figure~\ref{fig:RSR}
and are due to the coincidence of the NEOS and DANSS allowed regions
discussed above.
Therefore,
all the $3\sigma$
Neutrino-4 allowed regions are excluded at $2\sigma$ by the
reactor spectral-ratio and tritium measurements.

Moreover, the large-$\sin^2\!2\vartheta_{ee}$ parts of
the $3\sigma$ Neutrino-4 allowed regions
and most of
the $2\sigma$ Neutrino-4 allowed region
are excluded by the combined $3\sigma$
tritium and reactor spectral ratio bound.

\begin{figure}[!t]
\centering
\includegraphics*[width=\linewidth]{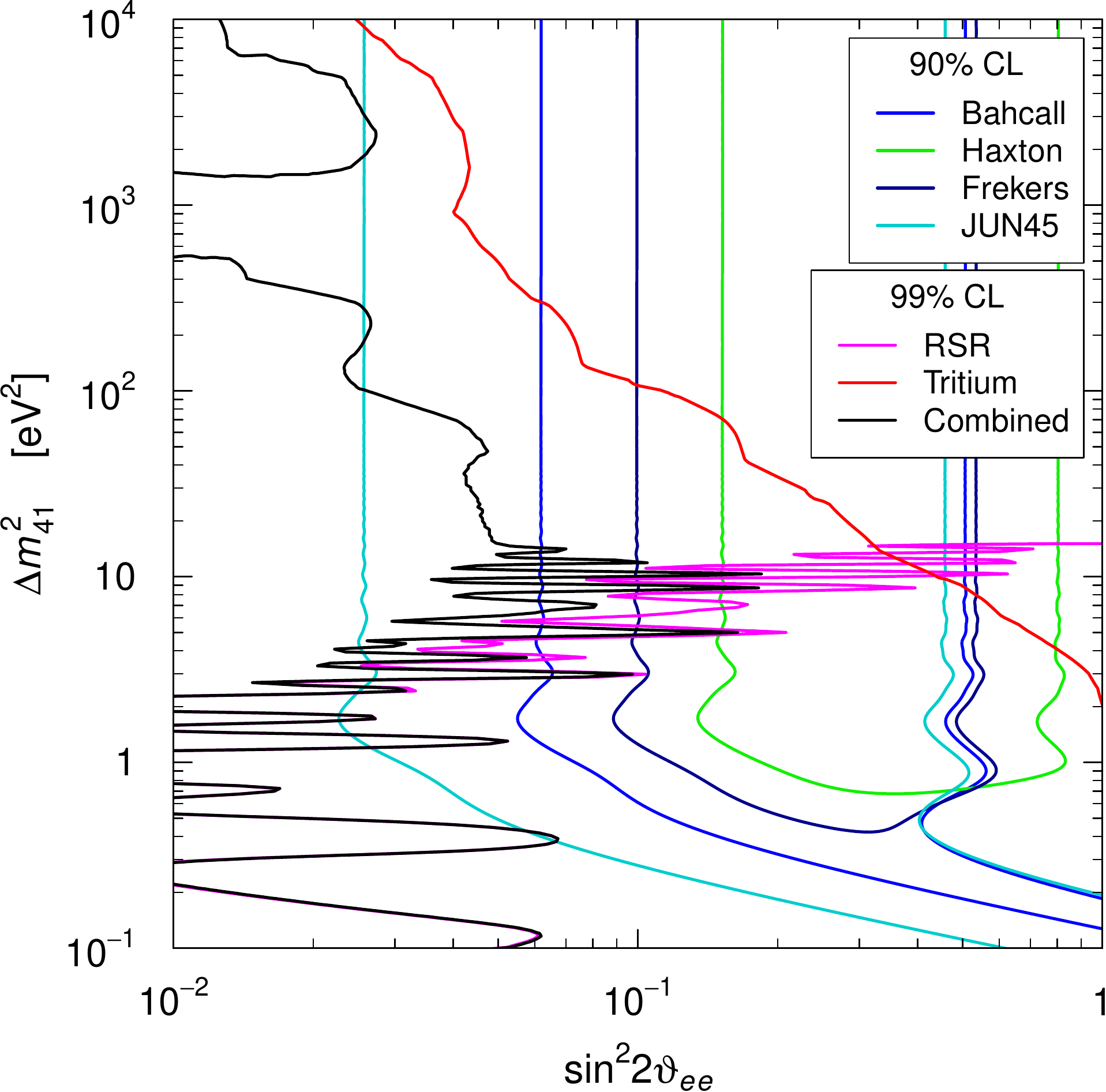}
\caption{ \label{fig:gallium}
Comparison of the regions in the
($\sin^2\!2\vartheta_{ee},\Delta{m}^2_{41}$)
plane allowed at 90\%~CL by the gallium neutrino anomaly
using the
Bahcall~\cite{Bahcall:1997eg},
Haxton~\cite{Haxton:1998uc},
Frekers~\cite{Frekers:2011zz}, and
JUN45~\cite{Kostensalo:2019vmv}
neutrino detection cross sections discussed in Ref.~\cite{Kostensalo:2019vmv}
with the
99\%~CL exclusion curves  obtained from the combined analysis of the data of
the Mainz, Troitsk and KATRIN tritium experiments
and the combined analysis of the reactor spectral ratio (RSR) measurements of the
Bugey-3, NEOS, PROSPECT and DANSS experiments.
Also shown is the combined tritium and reactor exclusion curve.
}
\end{figure}

\section{The gallium neutrino anomaly}
\label{sec:gallium}

Let us finally consider the gallium neutrino
anomaly~\cite{Abdurashitov:2005tb,Laveder:2007zz,Giunti:2006bj,Acero:2007su,Giunti:2009zz,Giunti:2010zu,Giunti:2012tn,Kostensalo:2019vmv},
that is a short-baseline disappearance of $\nu_{e}$'s
found in the gallium radioactive source experiments
GALLEX
\cite{Anselmann:1994ar,Hampel:1997fc,Kaether:2010ag}
and
SAGE
\cite{Abdurashitov:1996dp,Abdurashitov:1998ne,Abdurashitov:2005tb,Abdurashitov:2009tn}.
There is some uncertainty on the magnitude of the gallium neutrino anomaly,
that depends on the detection cross section,
which must be calculated, as in Refs.~\cite{Haxton:1998uc,Kostensalo:2019vmv},
or extrapolated from measurements of
$(p,n)$~\cite{Krofcheck:1985fg,Bahcall:1997eg}
or
$({}^{3}\text{He},{}^{3}\text{H})$~\cite{Frekers:2011zz}
charge-exchange reactions.
Figure~\ref{fig:gallium}
shows the regions in the
($\sin^2\!2\vartheta_{ee},\Delta{m}^2_{41}$)
plane allowed at 90\%~CL by the gallium neutrino anomaly
using the detection cross sections considered recently in Ref.~\cite{Kostensalo:2019vmv},
where a new shell model calculation based on
the effective Hamiltonian JUN45 was presented.
The Bahcall cross section was derived in Ref.~\cite{Bahcall:1997eg}
from the $(p,n)$ charge-exchange measurements in Ref.~\cite{Krofcheck:1985fg}.
The Haxton cross section was calculated in Ref.~\cite{Haxton:1998uc} using a shell model.
The Frekers cross section was obtained from the
$({}^{3}\text{He},{}^{3}\text{H})$~\cite{Frekers:2011zz}
charge-exchange measurements in Ref.~\cite{Frekers:2011zz}.

As done in Ref.~\cite{Kostensalo:2019vmv},
we show in Figure~\ref{fig:gallium} the contours of the 90\%~CL allowed regions
that have a lower bound for the effective mixing parameter $\sin^2\!2\vartheta_{ee}$.
One can see that
the relatively large Haxton cross section gives the strongest anomaly,
which requires rather large active sterile mixing
and is in severe tension with the tritium and reactor spectral ratio bounds.
Almost all the 90\%~CL Haxton allowed region is excluded at
99\%~CL by the combined tritium and reactor spectral ratio bound.
The smaller Frekers and Bahcall cross sections allow smaller values of the mixing,
but the corresponding 90\%~CL allowed regions in Figure~\ref{fig:gallium}
are in tension with the combined tritium and reactor spectral ratio 99\%~CL exclusion curve,
with only some very small not-excluded areas.
The JUN45 cross section is the smallest one and allows the smallest mixing,
as one can see from Figure~\ref{fig:gallium},
where the corresponding 90\%~CL allowed region has several areas that are not excluded by
the combined 99\%~CL tritium and reactor spectral ratio bound.
In particular, there is a large not-excluded area at large values of
$\Delta{m}^2_{41}$,
between about 5 and 100 $\text{eV}^2$.
These comparisons indicate that the smallest JUN45 gallium detection cross section
is favored with respect to the others.
Note that the KATRIN bound is essential for the exclusion of
large parts of the gallium allowed regions
for $\Delta{m}^2_{41}$ between about
$10$
and
$80 \, \text{eV}^2$,
where the KATRIN data dominate the tritium bound.

\section{Conclusions}
\label{sec:conclusions}

In this paper we have discussed the implications for 3+1 active-sterile neutrino mixing
of the recent KATRIN data~\cite{Aker:2019uuj}
on the search for the absolute value of neutrino masses.
We have first analyzed the KATRIN data in the framework of standard
three-neutrino mixing,
in order to check the validity of our method by comparing
the resulting bound on the effective mass $m_{\beta}$ with that obtained by the
KATRIN collaboration.
Then, we have presented the bounds obtained from the analysis of the KATRIN data
on the short-baseline oscillation parameters
$\sin^2\!2\vartheta_{ee}$ and $\Delta{m}^2_{41}$
in the framework of 3+1 active-sterile neutrino mixing.
We have shown that the KATRIN data allow to improve the bounds of the
Mainz~\cite{Kraus:2012he}
and
Troitsk~\cite{Belesev:2012hx,Belesev:2013cba}
experiments discussed in Ref.~\cite{Giunti:2012bc}
extending the excluded region from $\Delta{m}^2_{41} \approx 10-100 \, \text{eV}^2$
to
$\Delta{m}^2_{41} \approx 1-10 \, \text{eV}^2$
for large mixing
($\sin^2\!2\vartheta_{ee} \gtrsim 0.1$).
This result allows us to extend the exclusion of the large-$\Delta{m}^2_{41}$
solution of the Huber-Muller reactor antineutrino anomaly
to $\Delta{m}^2_{41} \approx 10 \, \text{eV}^2$
for
$\sin^2\!2\vartheta_{ee} \approx 0.1$
at 90\%~CL
(see Figure~\ref{fig:KMT}).

We also considered the model-independent bounds of the
Bugey-3~\cite{Declais:1995su},
NEOS~\cite{Ko:2016owz},
PROSPECT~\cite{Ashenfelter:2018iov}, and
DANSS~\cite{Alekseev:2018efk,Danilov:2019aef}
experiments that
measured the reactor antineutrino spectrum at different distances.
We have shown that there is a persistent model-independent
indication~\cite{Gariazzo:2018mwd,Dentler:2018sju,Berryman:2019hme}
of short-baseline oscillations
due to the coincidence of the NEOS and DANSS allowed regions,
albeit with a smaller statistical significance passing from the
2018~\cite{Alekseev:2018efk} to the 2019~\cite{Danilov:2019aef} DANSS data,
in agreement with the discussion in Ref.~\cite{Berryman:2019hme}.

The combination of the bounds of the reactor spectral ratio measurements
exclude most of the
low--$\Delta{m}^2_{41}$
solution of the Huber-Muller reactor antineutrino anomaly.
Therefore,
combining the tritium and reactor spectral ratio bounds,
we are able to exclude most of the region in the
($\sin^2\!2\vartheta_{ee},\Delta{m}^2_{41}$)
plane
corresponding to the
short-baseline solution of the Huber-Muller reactor antineutrino anomaly
(see Figure~\ref{fig:T+R}).

We also discussed the implications of these bounds for the interpretations of the
absolute reactor antineutrino rates assuming
one of the two recent new reactor neutrino flux calculations by
Estienne, Fallot et al. (EF)~\cite{Estienne:2019ujo}
and
Hayen, Kostensalo, Severijns, Suhonen (HKSS)~\cite{Hayen:2019eop}.
We have shown that the EF calculation,
that predicts a $^{235}\text{U}$ neutrino flux
that is smaller than that of Huber-Muller,
is in agreement with the bounds on 3+1 mixing
obtained from the tritium and reactor spectral ratio measurements.
On the other hand, since the HKSS calculation predicts reactor neutrino fluxes that are larger than
those of Huber-Muller,
the HKSS antineutrino anomaly region in the
($\sin^2\!2\vartheta_{ee},\Delta{m}^2_{41}$)
plane is more excluded than the Huber-Muller one
(see Figure~\ref{fig:EF-HKSS}).

We also compared the tritium and reactor spectral ratio bounds on 3+1 mixing
with the indication of large mixing of the Neutrino-4 reactor experiment~\cite{Serebrov:2018vdw}.
We have shown that the Neutrino-4 allowed regions in the
($\sin^2\!2\vartheta_{ee},\Delta{m}^2_{41}$)
plane are excluded at $2\sigma$ by the other reactor spectral ratio measurements.
The $3\sigma$ combined tritium and reactor spectral ratio bound
excludes the large-$\sin^2\!2\vartheta_{ee}$ parts of the
$3\sigma$ Neutrino-4 allowed regions.
(see Figure~\ref{fig:nu4}).

We finally considered the gallium neutrino anomaly
and we have shown that the combined bound of tritium and reactor spectral ratio measurements
favor the recent JUN45 shell model calculation of the neutrino-gallium cross section~\cite{Kostensalo:2019vmv}
with respect to older estimates~\cite{Bahcall:1997eg,Haxton:1998uc,Frekers:2011zz}.

\begin{acknowledgments}
We would like to thank Bryce Littlejohn,
Loredana Gastaldo and Francesco Vissani for useful discussions
that helped to improve the paper.
The work of C. Giunti was partially supported by the research grant "The Dark Universe: A Synergic Multimessenger Approach" number 2017X7X85K under the program PRIN 2017 funded by the Ministero dell'Istruzione, Universit\`a e della Ricerca (MIUR).
The work of Y.F. Li and Y.Y. Zhang was supported in part by Beijing Natural Science Foundation under Grant No.~1192019, and by the National Natural Science Foundation of China under Grant No.~11835013.
Y.F. Li is also grateful for the support by the CAS Center for Excellence in Particle Physics (CCEPP).
\end{acknowledgments}

%\bibliographystyle{apsrev4-1}
%\bibliography{katrin,new}

\begin{thebibliography}{65}%
\makeatletter
\providecommand \@ifxundefined [1]{%
\@ifx{#1\undefined}
}%
\providecommand \@ifnum [1]{%
\ifnum #1\expandafter \@firstoftwo
\else \expandafter \@secondoftwo
\fi
}%
\providecommand \@ifx [1]{%
\ifx #1\expandafter \@firstoftwo
\else \expandafter \@secondoftwo
\fi
}%
\providecommand \natexlab [1]{#1}%
\providecommand \enquote [1]{``#1''}%
\providecommand \bibnamefont [1]{#1}%
\providecommand \bibfnamefont [1]{#1}%
\providecommand \citenamefont [1]{#1}%
\providecommand \href@noop [0]{\@secondoftwo}%
\providecommand \href [0]{\begingroup \@sanitize@url \@href}%
\providecommand \@href[1]{\@@startlink{#1}\@@href}%
\providecommand \@@href[1]{\endgroup#1\@@endlink}%
\providecommand \@sanitize@url [0]{\catcode `\\12\catcode `\$12\catcode
`\&12\catcode `\#12\catcode `\^12\catcode `\_12\catcode `\%12\relax}%
\providecommand \@@startlink[1]{}%
\providecommand \@@endlink[0]{}%
\providecommand \url [0]{\begingroup\@sanitize@url \@url }%
\providecommand \@url [1]{\endgroup\@href {#1}{\urlprefix }}%
\providecommand \urlprefix [0]{URL }%
\providecommand \Eprint [0]{\href }%
\providecommand \doibase [0]{http://dx.doi.org/}%
\providecommand \selectlanguage [0]{\@gobble}%
\providecommand \bibinfo [0]{\@secondoftwo}%
\providecommand \bibfield [0]{\@secondoftwo}%
\providecommand \translation [1]{[#1]}%
\providecommand \BibitemOpen [0]{}%
\providecommand \bibitemStop [0]{}%
\providecommand \bibitemNoStop [0]{.\EOS\space}%
\providecommand \EOS [0]{\spacefactor3000\relax}%
\providecommand \BibitemShut [1]{\csname bibitem#1\endcsname}%
\let\auto@bib@innerbib\@empty
%</preamble>
\bibitem [{\citenamefont {Aker}\ \emph {et~al.}(2019)\citenamefont {Aker} \emph
{et~al.}}]{Aker:2019uuj}%
\BibitemOpen
\bibfield {author} {\bibinfo {author} {\bibfnamefont {M.}~\bibnamefont
{Aker}} \emph {et~al.} (\bibinfo {collaboration} {KATRIN}),\ }\href@noop {}
{\bibfield {journal} {\bibinfo {journal} {Phys.Rev.Lett.}\ }\textbf
{\bibinfo {volume} {123}},\ \bibinfo {pages} {221802} (\bibinfo {year}
{2019})},\ \Eprint {http://arxiv.org/abs/arXiv:1909.06048} {arXiv:1909.06048
[hep-ex]} \BibitemShut {NoStop}%
\bibitem [{\citenamefont {Giunti}\ and\ \citenamefont
{Lasserre}(2019)}]{Giunti:2019aiy}%
\BibitemOpen
\bibfield {author} {\bibinfo {author} {\bibfnamefont {C.}~\bibnamefont
{Giunti}}\ and\ \bibinfo {author} {\bibfnamefont {T.}~\bibnamefont
{Lasserre}},\ }\href {\doibase 10.1146/annurev-nucl-101918-023755} {\bibfield
{journal} {\bibinfo {journal} {Ann. Rev. Nucl. Part. Sci.}\ }\textbf
{\bibinfo {volume} {69}},\ \bibinfo {pages} {163} (\bibinfo {year} {2019})},\
\Eprint {http://arxiv.org/abs/arXiv:1901.08330} {arXiv:1901.08330 [hep-ph]}
\BibitemShut {NoStop}%
\bibitem [{\citenamefont {Diaz}\ \emph {et~al.}()\citenamefont {Diaz},
\citenamefont {Arguelles}, \citenamefont {Collin}, \citenamefont {Conrad},\
and\ \citenamefont {Shaevitz}}]{Diaz:2019fwt}%
\BibitemOpen
\bibfield {author} {\bibinfo {author} {\bibfnamefont {A.}~\bibnamefont
{Diaz}}, \bibinfo {author} {\bibfnamefont {C.}~\bibnamefont {Arguelles}},
\bibinfo {author} {\bibfnamefont {G.}~\bibnamefont {Collin}}, \bibinfo
{author} {\bibfnamefont {J.}~\bibnamefont {Conrad}}, \ and\ \bibinfo {author}
{\bibfnamefont {M.}~\bibnamefont {Shaevitz}},\ }\href@noop {} {\ }\Eprint
{http://arxiv.org/abs/arXiv:1906.00045} {arXiv:1906.00045 [hep-ex]}
\BibitemShut {NoStop}%
\bibitem [{\citenamefont {Boser}\ \emph {et~al.}(2020)\citenamefont {Boser},
\citenamefont {Buck}, \citenamefont {Giunti}, \citenamefont {Lesgourgues},
\citenamefont {Ludhova}, \citenamefont {Mertens}, \citenamefont {Schukraft},\
and\ \citenamefont {Wurm}}]{Boser:2019rta}%
\BibitemOpen
\bibfield {author} {\bibinfo {author} {\bibfnamefont {S.}~\bibnamefont
{Boser}}, \bibinfo {author} {\bibfnamefont {C.}~\bibnamefont {Buck}},
\bibinfo {author} {\bibfnamefont {C.}~\bibnamefont {Giunti}}, \bibinfo
{author} {\bibfnamefont {J.}~\bibnamefont {Lesgourgues}}, \bibinfo {author}
{\bibfnamefont {L.}~\bibnamefont {Ludhova}}, \bibinfo {author} {\bibfnamefont
{S.}~\bibnamefont {Mertens}}, \bibinfo {author} {\bibfnamefont
{A.}~\bibnamefont {Schukraft}}, \ and\ \bibinfo {author} {\bibfnamefont
{M.}~\bibnamefont {Wurm}},\ }\href@noop {} {\bibfield {journal} {\bibinfo
{journal} {Prog.Part.Nucl.Phys.}\ }\textbf {\bibinfo {volume} {111}},\
\bibinfo {pages} {103736} (\bibinfo {year} {2020})},\ \Eprint
{http://arxiv.org/abs/arXiv:1906.01739} {arXiv:1906.01739 [hep-ex]}
\BibitemShut {NoStop}%
\bibitem [{\citenamefont {Kraus}\ \emph {et~al.}(2013)\citenamefont {Kraus},
\citenamefont {Singer}, \citenamefont {Valerius},\ and\ \citenamefont
{Weinheimer}}]{Kraus:2012he}%
\BibitemOpen
\bibfield {author} {\bibinfo {author} {\bibfnamefont {C.}~\bibnamefont
{Kraus}}, \bibinfo {author} {\bibfnamefont {A.}~\bibnamefont {Singer}},
\bibinfo {author} {\bibfnamefont {K.}~\bibnamefont {Valerius}}, \ and\
\bibinfo {author} {\bibfnamefont {C.}~\bibnamefont {Weinheimer}},\ }\href
{\doibase 10.1140/epjc/s10052-013-2323-z} {\bibfield {journal} {\bibinfo
{journal} {Eur.Phys.J.}\ }\textbf {\bibinfo {volume} {C73}},\ \bibinfo
{pages} {2323} (\bibinfo {year} {2013})},\ \Eprint
{http://arxiv.org/abs/arXiv:1210.4194} {arXiv:1210.4194 [hep-ex]}
\BibitemShut {NoStop}%
\bibitem [{\citenamefont {Belesev}\ \emph {et~al.}(2013)\citenamefont
{Belesev}, \citenamefont {Berlev}, \citenamefont {Geraskin}, \citenamefont
{Golubev}, \citenamefont {Likhovid} \emph {et~al.}}]{Belesev:2012hx}%
\BibitemOpen
\bibfield {author} {\bibinfo {author} {\bibfnamefont {A.}~\bibnamefont
{Belesev}}, \bibinfo {author} {\bibfnamefont {A.}~\bibnamefont {Berlev}},
\bibinfo {author} {\bibfnamefont {E.}~\bibnamefont {Geraskin}}, \bibinfo
{author} {\bibfnamefont {A.}~\bibnamefont {Golubev}}, \bibinfo {author}
{\bibfnamefont {N.}~\bibnamefont {Likhovid}}, \emph {et~al.},\ }\href
{\doibase 10.1134/S0021364013020033} {\bibfield {journal} {\bibinfo
{journal} {JETP Lett.}\ }\textbf {\bibinfo {volume} {97}},\ \bibinfo {pages}
{67} (\bibinfo {year} {2013})},\ \Eprint
{http://arxiv.org/abs/arXiv:1211.7193} {arXiv:1211.7193 [hep-ex]}
\BibitemShut {NoStop}%
\bibitem [{\citenamefont {Belesev}\ \emph {et~al.}(2014)\citenamefont {Belesev}
\emph {et~al.}}]{Belesev:2013cba}%
\BibitemOpen
\bibfield {author} {\bibinfo {author} {\bibfnamefont {A.}~\bibnamefont
{Belesev}} \emph {et~al.},\ }\href@noop {} {\bibfield {journal} {\bibinfo
{journal} {J. Phys.}\ }\textbf {\bibinfo {volume} {G41}},\ \bibinfo {pages}
{015001} (\bibinfo {year} {2014})},\ \Eprint
{http://arxiv.org/abs/arXiv:1307.5687} {arXiv:1307.5687 [hep-ex]}
\BibitemShut {NoStop}%
\bibitem [{\citenamefont {Giunti}\ \emph {et~al.}(2013)\citenamefont {Giunti},
\citenamefont {Laveder}, \citenamefont {Li},\ and\ \citenamefont
{Long}}]{Giunti:2012bc}%
\BibitemOpen
\bibfield {author} {\bibinfo {author} {\bibfnamefont {C.}~\bibnamefont
{Giunti}}, \bibinfo {author} {\bibfnamefont {M.}~\bibnamefont {Laveder}},
\bibinfo {author} {\bibfnamefont {Y.~F.}\ \bibnamefont {Li}}, \ and\ \bibinfo
{author} {\bibfnamefont {H.}~\bibnamefont {Long}},\ }\href@noop {} {\bibfield
{journal} {\bibinfo {journal} {Phys. Rev.}\ }\textbf {\bibinfo {volume}
{D87}},\ \bibinfo {pages} {013004} (\bibinfo {year} {2013})},\ \Eprint
{http://arxiv.org/abs/arXiv:1212.3805} {arXiv:1212.3805 [hep-ph]}
\BibitemShut {NoStop}%
\bibitem [{\citenamefont {Mention}\ \emph {et~al.}(2011)\citenamefont {Mention}
\emph {et~al.}}]{Mention:2011rk}%
\BibitemOpen
\bibfield {author} {\bibinfo {author} {\bibfnamefont {G.}~\bibnamefont
{Mention}} \emph {et~al.},\ }\href@noop {} {\bibfield {journal} {\bibinfo
{journal} {Phys. Rev.}\ }\textbf {\bibinfo {volume} {D83}},\ \bibinfo {pages}
{073006} (\bibinfo {year} {2011})},\ \Eprint
{http://arxiv.org/abs/arXiv:1101.2755} {arXiv:1101.2755 [hep-ex]}
\BibitemShut {NoStop}%
\bibitem [{\citenamefont {Mueller}\ \emph {et~al.}(2011)\citenamefont {Mueller}
\emph {et~al.}}]{Mueller:2011nm}%
\BibitemOpen
\bibfield {author} {\bibinfo {author} {\bibfnamefont {T.~A.}\ \bibnamefont
{Mueller}} \emph {et~al.},\ }\href@noop {} {\bibfield {journal} {\bibinfo
{journal} {Phys. Rev.}\ }\textbf {\bibinfo {volume} {C83}},\ \bibinfo {pages}
{054615} (\bibinfo {year} {2011})},\ \Eprint
{http://arxiv.org/abs/arXiv:1101.2663} {arXiv:1101.2663 [hep-ex]}
\BibitemShut {NoStop}%
\bibitem [{\citenamefont {Huber}(2011)}]{Huber:2011wv}%
\BibitemOpen
\bibfield {author} {\bibinfo {author} {\bibfnamefont {P.}~\bibnamefont
{Huber}},\ }\href@noop {} {\bibfield {journal} {\bibinfo {journal} {Phys.
Rev.}\ }\textbf {\bibinfo {volume} {C84}},\ \bibinfo {pages} {024617}
(\bibinfo {year} {2011})},\ \Eprint {http://arxiv.org/abs/arXiv:1106.0687}
{arXiv:1106.0687 [hep-ph]} \BibitemShut {NoStop}%
\bibitem [{\citenamefont {Estienne}\ \emph {et~al.}(2019)\citenamefont
{Estienne}, \citenamefont {Fallot} \emph {et~al.}}]{Estienne:2019ujo}%
\BibitemOpen
\bibfield {author} {\bibinfo {author} {\bibfnamefont {M.}~\bibnamefont
{Estienne}}, \bibinfo {author} {\bibfnamefont {M.}~\bibnamefont {Fallot}},
\emph {et~al.},\ }\href {\doibase 10.1103/PhysRevLett.123.022502} {\bibfield
{journal} {\bibinfo {journal} {Phys. Rev. Lett.}\ }\textbf {\bibinfo
{volume} {123}},\ \bibinfo {pages} {022502} (\bibinfo {year} {2019})},\
\Eprint {http://arxiv.org/abs/arXiv:1904.09358} {arXiv:1904.09358 [nucl-ex]}
\BibitemShut {NoStop}%
\bibitem [{\citenamefont {Hayen}\ \emph {et~al.}(2019)\citenamefont {Hayen},
\citenamefont {Kostensalo}, \citenamefont {Severijns},\ and\ \citenamefont
{Suhonen}}]{Hayen:2019eop}%
\BibitemOpen
\bibfield {author} {\bibinfo {author} {\bibfnamefont {L.}~\bibnamefont
{Hayen}}, \bibinfo {author} {\bibfnamefont {J.}~\bibnamefont {Kostensalo}},
\bibinfo {author} {\bibfnamefont {N.}~\bibnamefont {Severijns}}, \ and\
\bibinfo {author} {\bibfnamefont {J.}~\bibnamefont {Suhonen}},\ }\href@noop
{} {\bibfield {journal} {\bibinfo {journal} {Phys.Rev.}\ }\textbf {\bibinfo
{volume} {C100}},\ \bibinfo {pages} {054323} (\bibinfo {year} {2019})},\
\Eprint {http://arxiv.org/abs/arXiv:1908.08302} {arXiv:1908.08302 [nucl-th]}
\BibitemShut {NoStop}%
\bibitem [{\citenamefont {Serebrov}\ \emph {et~al.}(2019)\citenamefont
{Serebrov} \emph {et~al.}}]{Serebrov:2018vdw}%
\BibitemOpen
\bibfield {author} {\bibinfo {author} {\bibfnamefont {A.}~\bibnamefont
{Serebrov}} \emph {et~al.} (\bibinfo {collaboration} {Neutrino-4}),\
}\href@noop {} {\bibfield {journal} {\bibinfo {journal} {Pisma
Zh.Eksp.Teor.Fiz.}\ }\textbf {\bibinfo {volume} {109}},\ \bibinfo {pages}
{209} (\bibinfo {year} {2019})},\ \Eprint
{http://arxiv.org/abs/arXiv:1809.10561} {arXiv:1809.10561 [hep-ex]}
\BibitemShut {NoStop}%
\bibitem [{\citenamefont {Konopinski}\ and\ \citenamefont
{Uhlenbeck}(1935)}]{Konopinski:1935zz}%
\BibitemOpen
\bibfield {author} {\bibinfo {author} {\bibfnamefont {E.~J.}\ \bibnamefont
{Konopinski}}\ and\ \bibinfo {author} {\bibfnamefont {G.~E.}\ \bibnamefont
{Uhlenbeck}},\ }\href {\doibase 10.1103/PhysRev.48.7} {\bibfield {journal}
{\bibinfo {journal} {Phys. Rev.}\ }\textbf {\bibinfo {volume} {48}},\
\bibinfo {pages} {7} (\bibinfo {year} {1935})}\BibitemShut {NoStop}%
\bibitem [{\citenamefont {Ludl}\ and\ \citenamefont
{Rodejohann}(2016)}]{Ludl:2016ane}%
\BibitemOpen
\bibfield {author} {\bibinfo {author} {\bibfnamefont {P.~O.}\ \bibnamefont
{Ludl}}\ and\ \bibinfo {author} {\bibfnamefont {W.}~\bibnamefont
{Rodejohann}},\ }\href@noop {} {\bibfield {journal} {\bibinfo {journal}
{JHEP}\ }\textbf {\bibinfo {volume} {1606}},\ \bibinfo {pages} {040}
(\bibinfo {year} {2016})},\ \Eprint {http://arxiv.org/abs/arXiv:1603.08690}
{arXiv:1603.08690 [hep-ph]} \BibitemShut {NoStop}%
\bibitem [{\citenamefont {Saenz}\ \emph {et~al.}(2000)\citenamefont {Saenz},
\citenamefont {Jonsell},\ and\ \citenamefont {Froelich}}]{Saenz:2000dul}%
\BibitemOpen
\bibfield {author} {\bibinfo {author} {\bibfnamefont {A.}~\bibnamefont
{Saenz}}, \bibinfo {author} {\bibfnamefont {S.}~\bibnamefont {Jonsell}}, \
and\ \bibinfo {author} {\bibfnamefont {P.}~\bibnamefont {Froelich}},\ }\href
{\doibase 10.1103/PhysRevLett.84.242} {\bibfield {journal} {\bibinfo
{journal} {Phys. Rev. Lett.}\ }\textbf {\bibinfo {volume} {84}},\ \bibinfo
{pages} {242} (\bibinfo {year} {2000})}\BibitemShut {NoStop}%
\bibitem [{\citenamefont {Doss}\ \emph {et~al.}(2006)\citenamefont {Doss},
\citenamefont {Tennyson}, \citenamefont {Saenz},\ and\ \citenamefont
{Jonsell}}]{Doss:2006zv}%
\BibitemOpen
\bibfield {author} {\bibinfo {author} {\bibfnamefont {N.}~\bibnamefont
{Doss}}, \bibinfo {author} {\bibfnamefont {J.}~\bibnamefont {Tennyson}},
\bibinfo {author} {\bibfnamefont {A.}~\bibnamefont {Saenz}}, \ and\ \bibinfo
{author} {\bibfnamefont {S.}~\bibnamefont {Jonsell}},\ }\href {\doibase
10.1103/PhysRevC.73.025502} {\bibfield {journal} {\bibinfo {journal} {Phys.
Rev.}\ }\textbf {\bibinfo {volume} {C73}},\ \bibinfo {pages} {025502}
(\bibinfo {year} {2006})}\BibitemShut {NoStop}%
\bibitem [{\citenamefont {Lobashev}\ and\ \citenamefont
{Spivak}(1985)}]{Lobashev:1985mu}%
\BibitemOpen
\bibfield {author} {\bibinfo {author} {\bibfnamefont {V.~M.}\ \bibnamefont
{Lobashev}}\ and\ \bibinfo {author} {\bibfnamefont {P.~E.}\ \bibnamefont
{Spivak}},\ }\href@noop {} {\bibfield {journal} {\bibinfo {journal} {Nucl.
Instrum. Meth.}\ }\textbf {\bibinfo {volume} {A240}},\ \bibinfo {pages} {305}
(\bibinfo {year} {1985})}\BibitemShut {NoStop}%
\bibitem [{\citenamefont {Picard}\ \emph {et~al.}(1992)\citenamefont {Picard},
\citenamefont {Backe}, \citenamefont {Barth}, \citenamefont {Bonn},
\citenamefont {Degen}, \citenamefont {Edling}, \citenamefont {Haid},
\citenamefont {Hermanni}, \citenamefont {Leiderer}, \citenamefont {Loeken},
\citenamefont {Molz}, \citenamefont {Moore}, \citenamefont {Osipowicz},
\citenamefont {Otten}, \citenamefont {Przyrembel}, \citenamefont {Schrader},
\citenamefont {Steininger},\ and\ \citenamefont
{Weinheimer}}]{PICARD1992345}%
\BibitemOpen
\bibfield {author} {\bibinfo {author} {\bibfnamefont {A.}~\bibnamefont
{Picard}}, \bibinfo {author} {\bibfnamefont {H.}~\bibnamefont {Backe}},
\bibinfo {author} {\bibfnamefont {H.}~\bibnamefont {Barth}}, \bibinfo
{author} {\bibfnamefont {J.}~\bibnamefont {Bonn}}, \bibinfo {author}
{\bibfnamefont {B.}~\bibnamefont {Degen}}, \bibinfo {author} {\bibfnamefont
{T.}~\bibnamefont {Edling}}, \bibinfo {author} {\bibfnamefont
{R.}~\bibnamefont {Haid}}, \bibinfo {author} {\bibfnamefont {A.}~\bibnamefont
{Hermanni}}, \bibinfo {author} {\bibfnamefont {P.}~\bibnamefont {Leiderer}},
\bibinfo {author} {\bibfnamefont {T.}~\bibnamefont {Loeken}}, \bibinfo
{author} {\bibfnamefont {A.}~\bibnamefont {Molz}}, \bibinfo {author}
{\bibfnamefont {R.}~\bibnamefont {Moore}}, \bibinfo {author} {\bibfnamefont
{A.}~\bibnamefont {Osipowicz}}, \bibinfo {author} {\bibfnamefont
{E.}~\bibnamefont {Otten}}, \bibinfo {author} {\bibfnamefont
{M.}~\bibnamefont {Przyrembel}}, \bibinfo {author} {\bibfnamefont
{M.}~\bibnamefont {Schrader}}, \bibinfo {author} {\bibfnamefont
{M.}~\bibnamefont {Steininger}}, \ and\ \bibinfo {author} {\bibfnamefont
{C.}~\bibnamefont {Weinheimer}},\ }\href {\doibase
https://doi.org/10.1016/0168-583X(92)95119-C} {\bibfield {journal} {\bibinfo
{journal} {Nuclear Instruments and Methods in Physics Research Section B:
Beam Interactions with Materials and Atoms}\ }\textbf {\bibinfo {volume}
{63}},\ \bibinfo {pages} {345 } (\bibinfo {year} {1992})}\BibitemShut
{NoStop}%
\bibitem [{\citenamefont {Kleesiek}\ \emph {et~al.}(2019)\citenamefont
{Kleesiek} \emph {et~al.}}]{Kleesiek:2018mel}%
\BibitemOpen
\bibfield {author} {\bibinfo {author} {\bibfnamefont {M.}~\bibnamefont
{Kleesiek}} \emph {et~al.},\ }\href@noop {} {\bibfield {journal} {\bibinfo
{journal} {Eur.Phys.J.}\ }\textbf {\bibinfo {volume} {C79}},\ \bibinfo
{pages} {204} (\bibinfo {year} {2019})},\ \Eprint
{http://arxiv.org/abs/arXiv:1806.00369} {arXiv:1806.00369 [physics]}
\BibitemShut {NoStop}%
\bibitem [{\citenamefont {Feldman}\ and\ \citenamefont
{Cousins}(1998)}]{Feldman:1997qc}%
\BibitemOpen
\bibfield {author} {\bibinfo {author} {\bibfnamefont {G.~J.}\ \bibnamefont
{Feldman}}\ and\ \bibinfo {author} {\bibfnamefont {R.~D.}\ \bibnamefont
{Cousins}},\ }\href@noop {} {\bibfield {journal} {\bibinfo {journal} {Phys.
Rev.}\ }\textbf {\bibinfo {volume} {D57}},\ \bibinfo {pages} {3873} (\bibinfo
{year} {1998})},\ \Eprint {http://arxiv.org/abs/physics/9711021}
{physics/9711021} \BibitemShut {NoStop}%
\bibitem [{\citenamefont {de~Salas}\ \emph
{et~al.}(2018{\natexlab{a}})\citenamefont {de~Salas}, \citenamefont {Forero},
\citenamefont {Ternes}, \citenamefont {Tortola},\ and\ \citenamefont
{Valle}}]{deSalas:2017kay}%
\BibitemOpen
\bibfield {author} {\bibinfo {author} {\bibfnamefont {P.~F.}\ \bibnamefont
{de~Salas}}, \bibinfo {author} {\bibfnamefont {D.~V.}\ \bibnamefont
{Forero}}, \bibinfo {author} {\bibfnamefont {C.~A.}\ \bibnamefont {Ternes}},
\bibinfo {author} {\bibfnamefont {M.}~\bibnamefont {Tortola}}, \ and\
\bibinfo {author} {\bibfnamefont {J.~W.~F.}\ \bibnamefont {Valle}},\
}\href@noop {} {\bibfield {journal} {\bibinfo {journal} {Phys.Lett.}\
}\textbf {\bibinfo {volume} {B782}},\ \bibinfo {pages} {633} (\bibinfo {year}
{2018}{\natexlab{a}})},\ \Eprint {http://arxiv.org/abs/arXiv:1708.01186}
{arXiv:1708.01186 [hep-ph]} \BibitemShut {NoStop}%
\bibitem [{\citenamefont {Capozzi}\ \emph {et~al.}(2018)\citenamefont
{Capozzi}, \citenamefont {Lisi}, \citenamefont {Marrone},\ and\ \citenamefont
{Palazzo}}]{Capozzi:2018ubv}%
\BibitemOpen
\bibfield {author} {\bibinfo {author} {\bibfnamefont {F.}~\bibnamefont
{Capozzi}}, \bibinfo {author} {\bibfnamefont {E.}~\bibnamefont {Lisi}},
\bibinfo {author} {\bibfnamefont {A.}~\bibnamefont {Marrone}}, \ and\
\bibinfo {author} {\bibfnamefont {A.}~\bibnamefont {Palazzo}},\ }\href@noop
{} {\bibfield {journal} {\bibinfo {journal} {Prog.Part.Nucl.Phys.}\
}\textbf {\bibinfo {volume} {102}},\ \bibinfo {pages} {48} (\bibinfo {year}
{2018})},\ \Eprint {http://arxiv.org/abs/arXiv:1804.09678} {arXiv:1804.09678
[hep-ph]} \BibitemShut {NoStop}%
\bibitem [{\citenamefont {Esteban}\ \emph {et~al.}(2019)\citenamefont
{Esteban}, \citenamefont {Gonzalez-Garcia}, \citenamefont
{Hernandez-Cabezudo}, \citenamefont {Maltoni},\ and\ \citenamefont
{Schwetz}}]{Esteban:2018azc}%
\BibitemOpen
\bibfield {author} {\bibinfo {author} {\bibfnamefont {I.}~\bibnamefont
{Esteban}}, \bibinfo {author} {\bibfnamefont {M.}~\bibnamefont
{Gonzalez-Garcia}}, \bibinfo {author} {\bibfnamefont {A.}~\bibnamefont
{Hernandez-Cabezudo}}, \bibinfo {author} {\bibfnamefont {M.}~\bibnamefont
{Maltoni}}, \ and\ \bibinfo {author} {\bibfnamefont {T.}~\bibnamefont
{Schwetz}},\ }\href@noop {} {\bibfield {journal} {\bibinfo {journal}
{JHEP}\ }\textbf {\bibinfo {volume} {1901}},\ \bibinfo {pages} {106}
(\bibinfo {year} {2019})},\ \Eprint {http://arxiv.org/abs/arXiv:1811.05487}
{arXiv:1811.05487 [hep-ph]} \BibitemShut {NoStop}%
\bibitem [{\citenamefont {Tanabashi}\ \emph {et~al.}(2018)\citenamefont
{Tanabashi} \emph {et~al.}}]{Tanabashi:2018oca}%
\BibitemOpen
\bibfield {author} {\bibinfo {author} {\bibfnamefont {M.}~\bibnamefont
{Tanabashi}} \emph {et~al.} (\bibinfo {collaboration} {Particle Data
Group}),\ }\href {\doibase 10.1103/PhysRevD.98.030001} {\bibfield {journal}
{\bibinfo {journal} {Phys. Rev.}\ }\textbf {\bibinfo {volume} {D98}},\
\bibinfo {pages} {030001} (\bibinfo {year} {2018})}\BibitemShut {NoStop}%
\bibitem [{\citenamefont {de~Salas}\ \emph
{et~al.}(2018{\natexlab{b}})\citenamefont {de~Salas}, \citenamefont
{Gariazzo}, \citenamefont {Mena}, \citenamefont {Ternes},\ and\ \citenamefont
{Tortola}}]{deSalas:2018bym}%
\BibitemOpen
\bibfield {author} {\bibinfo {author} {\bibfnamefont {P.~F.}\ \bibnamefont
{de~Salas}}, \bibinfo {author} {\bibfnamefont {S.}~\bibnamefont {Gariazzo}},
\bibinfo {author} {\bibfnamefont {O.}~\bibnamefont {Mena}}, \bibinfo {author}
{\bibfnamefont {C.~A.}\ \bibnamefont {Ternes}}, \ and\ \bibinfo {author}
{\bibfnamefont {M.}~\bibnamefont {Tortola}},\ }\href@noop {} {\bibfield
{journal} {\bibinfo {journal} {Front.Astron.Space Sci.}\ }\textbf {\bibinfo
{volume} {5}},\ \bibinfo {pages} {36} (\bibinfo {year}
{2018}{\natexlab{b}})},\ \Eprint {http://arxiv.org/abs/arXiv:1806.11051}
{arXiv:1806.11051 [hep-ph]} \BibitemShut {NoStop}%
\bibitem [{\citenamefont {Giunti}\ \emph {et~al.}(2019)\citenamefont {Giunti},
\citenamefont {Li}, \citenamefont {Littlejohn},\ and\ \citenamefont
{Surukuchi}}]{Giunti:2019qlt}%
\BibitemOpen
\bibfield {author} {\bibinfo {author} {\bibfnamefont {C.}~\bibnamefont
{Giunti}}, \bibinfo {author} {\bibfnamefont {Y.~F.}\ \bibnamefont {Li}},
\bibinfo {author} {\bibfnamefont {B.~R.}\ \bibnamefont {Littlejohn}}, \ and\
\bibinfo {author} {\bibfnamefont {P.~T.}\ \bibnamefont {Surukuchi}},\
}\href@noop {} {\bibfield {journal} {\bibinfo {journal} {Phys.Rev.}\
}\textbf {\bibinfo {volume} {D99}},\ \bibinfo {pages} {073005} (\bibinfo
{year} {2019})},\ \Eprint {http://arxiv.org/abs/arXiv:1901.01807}
{arXiv:1901.01807 [hep-ph]} \BibitemShut {NoStop}%
\bibitem [{\citenamefont {Berryman}\ and\ \citenamefont
{Huber}(2020)}]{Berryman:2019hme}%
\BibitemOpen
\bibfield {author} {\bibinfo {author} {\bibfnamefont {J.}~\bibnamefont
{Berryman}}\ and\ \bibinfo {author} {\bibfnamefont {P.}~\bibnamefont
{Huber}},\ }\href@noop {} {\bibfield {journal} {\bibinfo {journal}
{Phys.Rev.}\ }\textbf {\bibinfo {volume} {D101}},\ \bibinfo {pages} {015008}
(\bibinfo {year} {2020})},\ \Eprint {http://arxiv.org/abs/arXiv:1909.09267}
{arXiv:1909.09267 [hep-ph]} \BibitemShut {NoStop}%
\bibitem [{\citenamefont {Giunti}\ \emph {et~al.}(2012)\citenamefont {Giunti},
\citenamefont {Laveder}, \citenamefont {Li}, \citenamefont {Liu},\ and\
\citenamefont {Long}}]{Giunti:2012tn}%
\BibitemOpen
\bibfield {author} {\bibinfo {author} {\bibfnamefont {C.}~\bibnamefont
{Giunti}}, \bibinfo {author} {\bibfnamefont {M.}~\bibnamefont {Laveder}},
\bibinfo {author} {\bibfnamefont {Y.~F.}\ \bibnamefont {Li}}, \bibinfo
{author} {\bibfnamefont {Q.}~\bibnamefont {Liu}}, \ and\ \bibinfo {author}
{\bibfnamefont {H.}~\bibnamefont {Long}},\ }\href@noop {} {\bibfield
{journal} {\bibinfo {journal} {Phys. Rev.}\ }\textbf {\bibinfo {volume}
{D86}},\ \bibinfo {pages} {113014} (\bibinfo {year} {2012})},\ \Eprint
{http://arxiv.org/abs/arXiv:1210.5715} {arXiv:1210.5715 [hep-ph]}
\BibitemShut {NoStop}%
\bibitem [{\citenamefont {Achkar}\ \emph {et~al.}(1995)\citenamefont {Achkar}
\emph {et~al.}}]{Declais:1995su}%
\BibitemOpen
\bibfield {author} {\bibinfo {author} {\bibfnamefont {B.}~\bibnamefont
{Achkar}} \emph {et~al.} (\bibinfo {collaboration} {Bugey}),\ }\href@noop {}
{\bibfield {journal} {\bibinfo {journal} {Nucl. Phys.}\ }\textbf {\bibinfo
{volume} {B434}},\ \bibinfo {pages} {503} (\bibinfo {year}
{1995})}\BibitemShut {NoStop}%
\bibitem [{\citenamefont {Kopp}\ \emph {et~al.}(2013)\citenamefont {Kopp},
\citenamefont {Machado}, \citenamefont {Maltoni},\ and\ \citenamefont
{Schwetz}}]{Kopp:2013vaa}%
\BibitemOpen
\bibfield {author} {\bibinfo {author} {\bibfnamefont {J.}~\bibnamefont
{Kopp}}, \bibinfo {author} {\bibfnamefont {P.~A.~N.}\ \bibnamefont
{Machado}}, \bibinfo {author} {\bibfnamefont {M.}~\bibnamefont {Maltoni}}, \
and\ \bibinfo {author} {\bibfnamefont {T.}~\bibnamefont {Schwetz}},\ }\href
{\doibase 10.1007/JHEP05(2013)050} {\bibfield {journal} {\bibinfo {journal}
{JHEP}\ }\textbf {\bibinfo {volume} {1305}},\ \bibinfo {pages} {050}
(\bibinfo {year} {2013})},\ \Eprint {http://arxiv.org/abs/arXiv:1303.3011}
{arXiv:1303.3011 [hep-ph]} \BibitemShut {NoStop}%
\bibitem [{\citenamefont {Alekseev}\ \emph {et~al.}(2018)\citenamefont
{Alekseev} \emph {et~al.}}]{Alekseev:2018efk}%
\BibitemOpen
\bibfield {author} {\bibinfo {author} {\bibfnamefont {I.}~\bibnamefont
{Alekseev}} \emph {et~al.} (\bibinfo {collaboration} {DANSS}),\ }\href@noop
{} {\bibfield {journal} {\bibinfo {journal} {Phys.Lett.}\ }\textbf
{\bibinfo {volume} {B787}},\ \bibinfo {pages} {56} (\bibinfo {year}
{2018})},\ \Eprint {http://arxiv.org/abs/arXiv:1804.04046} {arXiv:1804.04046
[hep-ex]} \BibitemShut {NoStop}%
\bibitem [{\citenamefont {Danilov}()}]{Danilov:2019aef}%
\BibitemOpen
\bibfield {author} {\bibinfo {author} {\bibfnamefont {M.}~\bibnamefont
{Danilov}} (\bibinfo {collaboration} {DANSS}),\ }\href@noop {} {\ }\Eprint
{http://arxiv.org/abs/arXiv:1911.10140} {arXiv:1911.10140 [hep-ex]}
\BibitemShut {NoStop}%
\bibitem [{\citenamefont {Ashenfelter}\ \emph {et~al.}(2018)\citenamefont
{Ashenfelter} \emph {et~al.}}]{Ashenfelter:2018iov}%
\BibitemOpen
\bibfield {author} {\bibinfo {author} {\bibfnamefont {J.}~\bibnamefont
{Ashenfelter}} \emph {et~al.} (\bibinfo {collaboration} {PROSPECT}),\
}\href@noop {} {\bibfield {journal} {\bibinfo {journal} {Phys.Rev.Lett.}\
}\textbf {\bibinfo {volume} {121}},\ \bibinfo {pages} {251802} (\bibinfo
{year} {2018})},\ \Eprint {http://arxiv.org/abs/arXiv:1806.02784}
{arXiv:1806.02784 [hep-ex]} \BibitemShut {NoStop}%
\bibitem [{\citenamefont {Almazan}\ \emph {et~al.}(2018)\citenamefont {Almazan}
\emph {et~al.}}]{Almazan:2018wln}%
\BibitemOpen
\bibfield {author} {\bibinfo {author} {\bibfnamefont {H.}~\bibnamefont
{Almazan}} \emph {et~al.} (\bibinfo {collaboration} {STEREO}),\ }\href@noop
{} {\bibfield {journal} {\bibinfo {journal} {Phys.Rev.Lett.}\ }\textbf
{\bibinfo {volume} {121}},\ \bibinfo {pages} {161801} (\bibinfo {year}
{2018})},\ \Eprint {http://arxiv.org/abs/arXiv:1806.02096} {arXiv:1806.02096
[hep-ex]} \BibitemShut {NoStop}%
\bibitem [{\citenamefont {Almazan~Molina}\ \emph {et~al.}()\citenamefont
{Almazan~Molina} \emph {et~al.}}]{AlmazanMolina:2019qul}%
\BibitemOpen
\bibfield {author} {\bibinfo {author} {\bibfnamefont {H.}~\bibnamefont
{Almazan~Molina}} \emph {et~al.} (\bibinfo {collaboration} {STEREO}),\
}\href@noop {} {\ }\Eprint {http://arxiv.org/abs/arXiv:1912.06582}
{arXiv:1912.06582 [hep-ex]} \BibitemShut {NoStop}%
\bibitem [{\citenamefont {Ko}\ \emph {et~al.}(2017)\citenamefont {Ko} \emph
{et~al.}}]{Ko:2016owz}%
\BibitemOpen
\bibfield {author} {\bibinfo {author} {\bibfnamefont {Y.}~\bibnamefont {Ko}}
\emph {et~al.} (\bibinfo {collaboration} {NEOS}),\ }\href@noop {} {\bibfield
{journal} {\bibinfo {journal} {Phys.Rev.Lett.}\ }\textbf {\bibinfo {volume}
{118}},\ \bibinfo {pages} {121802} (\bibinfo {year} {2017})},\ \Eprint
{http://arxiv.org/abs/arXiv:1610.05134} {arXiv:1610.05134 [hep-ex]}
\BibitemShut {NoStop}%
\bibitem [{\citenamefont {An}\ \emph {et~al.}(2017)\citenamefont {An} \emph
{et~al.}}]{An:2016srz}%
\BibitemOpen
\bibfield {author} {\bibinfo {author} {\bibfnamefont {F.}~\bibnamefont {An}}
\emph {et~al.} (\bibinfo {collaboration} {Daya Bay}),\ }\href@noop {}
{\bibfield {journal} {\bibinfo {journal} {Chin.Phys.}\ }\textbf {\bibinfo
{volume} {C41}},\ \bibinfo {pages} {013002} (\bibinfo {year} {2017})},\
\Eprint {http://arxiv.org/abs/arXiv:1607.05378} {arXiv:1607.05378 [hep-ex]}
\BibitemShut {NoStop}%
\bibitem [{\citenamefont {Gariazzo}\ \emph {et~al.}(2017)\citenamefont
{Gariazzo}, \citenamefont {Giunti}, \citenamefont {Laveder},\ and\
\citenamefont {Li}}]{Gariazzo:2017fdh}%
\BibitemOpen
\bibfield {author} {\bibinfo {author} {\bibfnamefont {S.}~\bibnamefont
{Gariazzo}}, \bibinfo {author} {\bibfnamefont {C.}~\bibnamefont {Giunti}},
\bibinfo {author} {\bibfnamefont {M.}~\bibnamefont {Laveder}}, \ and\
\bibinfo {author} {\bibfnamefont {Y.~F.}\ \bibnamefont {Li}},\ }\href@noop {}
{\bibfield {journal} {\bibinfo {journal} {JHEP}\ }\textbf {\bibinfo
{volume} {1706}},\ \bibinfo {pages} {135} (\bibinfo {year} {2017})},\ \Eprint
{http://arxiv.org/abs/arXiv:1703.00860} {arXiv:1703.00860 [hep-ph]}
\BibitemShut {NoStop}%
\bibitem [{\citenamefont {Gariazzo}\ \emph {et~al.}(2018)\citenamefont
{Gariazzo}, \citenamefont {Giunti}, \citenamefont {Laveder},\ and\
\citenamefont {Li}}]{Gariazzo:2018mwd}%
\BibitemOpen
\bibfield {author} {\bibinfo {author} {\bibfnamefont {S.}~\bibnamefont
{Gariazzo}}, \bibinfo {author} {\bibfnamefont {C.}~\bibnamefont {Giunti}},
\bibinfo {author} {\bibfnamefont {M.}~\bibnamefont {Laveder}}, \ and\
\bibinfo {author} {\bibfnamefont {Y.~F.}\ \bibnamefont {Li}},\ }\href@noop {}
{\bibfield {journal} {\bibinfo {journal} {Phys.Lett.}\ }\textbf {\bibinfo
{volume} {B782}},\ \bibinfo {pages} {13} (\bibinfo {year} {2018})},\ \Eprint
{http://arxiv.org/abs/arXiv:1801.06467} {arXiv:1801.06467 [hep-ph]}
\BibitemShut {NoStop}%
\bibitem [{\citenamefont {Dentler}\ \emph {et~al.}(2018)\citenamefont
{Dentler}, \citenamefont {Hernandez-Cabezudo}, \citenamefont {Kopp},
\citenamefont {Machado}, \citenamefont {Maltoni}, \citenamefont
{Martinez-Soler},\ and\ \citenamefont {Schwetz}}]{Dentler:2018sju}%
\BibitemOpen
\bibfield {author} {\bibinfo {author} {\bibfnamefont {M.}~\bibnamefont
{Dentler}}, \bibinfo {author} {\bibfnamefont {A.}~\bibnamefont
{Hernandez-Cabezudo}}, \bibinfo {author} {\bibfnamefont {J.}~\bibnamefont
{Kopp}}, \bibinfo {author} {\bibfnamefont {P.~A.~N.}\ \bibnamefont
{Machado}}, \bibinfo {author} {\bibfnamefont {M.}~\bibnamefont {Maltoni}},
\bibinfo {author} {\bibfnamefont {I.}~\bibnamefont {Martinez-Soler}}, \ and\
\bibinfo {author} {\bibfnamefont {T.}~\bibnamefont {Schwetz}},\ }\href@noop
{} {\bibfield {journal} {\bibinfo {journal} {JHEP}\ }\textbf {\bibinfo
{volume} {1808}},\ \bibinfo {pages} {010} (\bibinfo {year} {2018})},\ \Eprint
{http://arxiv.org/abs/arXiv:1803.10661} {arXiv:1803.10661 [hep-ph]}
\BibitemShut {NoStop}%
\bibitem [{\citenamefont {Seo}(2015)}]{Seo:2014xei}%
\BibitemOpen
\bibfield {author} {\bibinfo {author} {\bibfnamefont {S.-H.}\ \bibnamefont
{Seo}} (\bibinfo {collaboration} {RENO}),\ }\href@noop {} {\bibfield
{journal} {\bibinfo {journal} {AIP Conf. Proc.}\ }\textbf {\bibinfo {volume}
{1666}},\ \bibinfo {pages} {080002} (\bibinfo {year} {2015})},\ \Eprint
{http://arxiv.org/abs/arXiv:1410.7987} {arXiv:1410.7987 [hep-ex]}
\BibitemShut {NoStop}%
\bibitem [{\citenamefont {Choi}\ \emph {et~al.}(2016)\citenamefont {Choi} \emph
{et~al.}}]{RENO:2015ksa}%
\BibitemOpen
\bibfield {author} {\bibinfo {author} {\bibfnamefont {J.}~\bibnamefont
{Choi}} \emph {et~al.} (\bibinfo {collaboration} {RENO}),\ }\href@noop {}
{\bibfield {journal} {\bibinfo {journal} {Phys. Rev. Lett.}\ }\textbf
{\bibinfo {volume} {116}},\ \bibinfo {pages} {211801} (\bibinfo {year}
{2016})},\ \Eprint {http://arxiv.org/abs/arXiv:1511.05849} {arXiv:1511.05849
[hep-ex]} \BibitemShut {NoStop}%
\bibitem [{\citenamefont {Abe}\ \emph {et~al.}(2014)\citenamefont {Abe} \emph
{et~al.}}]{Abe:2014bwa}%
\BibitemOpen
\bibfield {author} {\bibinfo {author} {\bibfnamefont {Y.}~\bibnamefont
{Abe}} \emph {et~al.} (\bibinfo {collaboration} {Double Chooz}),\ }\href
{\doibase 10.1007/JHEP02(2015)074, 10.1007/JHEP10(2014)086} {\bibfield
{journal} {\bibinfo {journal} {JHEP}\ }\textbf {\bibinfo {volume} {10}},\
\bibinfo {pages} {086} (\bibinfo {year} {2014})},\ \bibinfo {note} {[Erratum:
JHEP 02, 074 (2015)]},\ \Eprint {http://arxiv.org/abs/arXiv:1406.7763}
{arXiv:1406.7763 [hep-ex]} \BibitemShut {NoStop}%
\bibitem [{\citenamefont {Huber}(2016)}]{Huber:2016fkt}%
\BibitemOpen
\bibfield {author} {\bibinfo {author} {\bibfnamefont {P.}~\bibnamefont
{Huber}},\ }\href@noop {} {\bibfield {journal} {\bibinfo {journal} {Nucl.
Phys.}\ }\textbf {\bibinfo {volume} {B908}},\ \bibinfo {pages} {268}
(\bibinfo {year} {2016})},\ \Eprint {http://arxiv.org/abs/arXiv:1602.01499}
{arXiv:1602.01499 [hep-ph]} \BibitemShut {NoStop}%
\bibitem [{\citenamefont {Hayes}\ and\ \citenamefont
{Vogel}(2016)}]{Hayes:2016qnu}%
\BibitemOpen
\bibfield {author} {\bibinfo {author} {\bibfnamefont {A.~C.}\ \bibnamefont
{Hayes}}\ and\ \bibinfo {author} {\bibfnamefont {P.}~\bibnamefont {Vogel}},\
}\href@noop {} {\bibfield {journal} {\bibinfo {journal}
{Ann.Rev.Nucl.Part.Sci.}\ }\textbf {\bibinfo {volume} {66}},\ \bibinfo
{pages} {219} (\bibinfo {year} {2016})},\ \Eprint
{http://arxiv.org/abs/arXiv:1605.02047} {arXiv:1605.02047 [hep-ph]}
\BibitemShut {NoStop}%
\bibitem [{\citenamefont {Hayes}\ \emph {et~al.}(2018)\citenamefont {Hayes}
\emph {et~al.}}]{Hayes:2017res}%
\BibitemOpen
\bibfield {author} {\bibinfo {author} {\bibfnamefont {A.}~\bibnamefont
{Hayes}} \emph {et~al.},\ }\href@noop {} {\bibfield {journal} {\bibinfo
{journal} {Phys.Rev.Lett.}\ }\textbf {\bibinfo {volume} {120}},\ \bibinfo
{pages} {022503} (\bibinfo {year} {2018})},\ \Eprint
{http://arxiv.org/abs/arXiv:1707.07728} {arXiv:1707.07728 [nucl-th]}
\BibitemShut {NoStop}%
\bibitem [{\citenamefont {Bahcall}(1997)}]{Bahcall:1997eg}%
\BibitemOpen
\bibfield {author} {\bibinfo {author} {\bibfnamefont {J.~N.}\ \bibnamefont
{Bahcall}},\ }\href@noop {} {\bibfield {journal} {\bibinfo {journal} {Phys.
Rev.}\ }\textbf {\bibinfo {volume} {C56}},\ \bibinfo {pages} {3391} (\bibinfo
{year} {1997})},\ \Eprint {http://arxiv.org/abs/hep-ph/9710491}
{hep-ph/9710491} \BibitemShut {NoStop}%
\bibitem [{\citenamefont {Haxton}(1998)}]{Haxton:1998uc}%
\BibitemOpen
\bibfield {author} {\bibinfo {author} {\bibfnamefont {W.~C.}\ \bibnamefont
{Haxton}},\ }\href {\doibase 10.1016/S0370-2693(98)00581-4} {\bibfield
{journal} {\bibinfo {journal} {Phys. Lett.}\ }\textbf {\bibinfo {volume}
{B431}},\ \bibinfo {pages} {110} (\bibinfo {year} {1998})},\ \Eprint
{http://arxiv.org/abs/nucl-th/9804011} {nucl-th/9804011} \BibitemShut
{NoStop}%
\bibitem [{\citenamefont {Frekers}\ \emph {et~al.}(2011)\citenamefont
{Frekers}, \citenamefont {Ejiri}, \citenamefont {Akimune}, \citenamefont
{Adachi}, \citenamefont {Bilgier} \emph {et~al.}}]{Frekers:2011zz}%
\BibitemOpen
\bibfield {author} {\bibinfo {author} {\bibfnamefont {D.}~\bibnamefont
{Frekers}}, \bibinfo {author} {\bibfnamefont {H.}~\bibnamefont {Ejiri}},
\bibinfo {author} {\bibfnamefont {H.}~\bibnamefont {Akimune}}, \bibinfo
{author} {\bibfnamefont {T.}~\bibnamefont {Adachi}}, \bibinfo {author}
{\bibfnamefont {B.}~\bibnamefont {Bilgier}}, \emph {et~al.},\ }\href@noop {}
{\bibfield {journal} {\bibinfo {journal} {Phys. Lett.}\ }\textbf {\bibinfo
{volume} {B706}},\ \bibinfo {pages} {134} (\bibinfo {year}
{2011})}\BibitemShut {NoStop}%
\bibitem [{\citenamefont {Kostensalo}\ \emph {et~al.}(2019)\citenamefont
{Kostensalo}, \citenamefont {Suhonen}, \citenamefont {Giunti},\ and\
\citenamefont {Srivastava}}]{Kostensalo:2019vmv}%
\BibitemOpen
\bibfield {author} {\bibinfo {author} {\bibfnamefont {J.}~\bibnamefont
{Kostensalo}}, \bibinfo {author} {\bibfnamefont {J.}~\bibnamefont {Suhonen}},
\bibinfo {author} {\bibfnamefont {C.}~\bibnamefont {Giunti}}, \ and\ \bibinfo
{author} {\bibfnamefont {P.~C.}\ \bibnamefont {Srivastava}},\ }\href@noop {}
{\bibfield {journal} {\bibinfo {journal} {Phys.Lett.}\ }\textbf {\bibinfo
{volume} {B795}},\ \bibinfo {pages} {542} (\bibinfo {year} {2019})},\ \Eprint
{http://arxiv.org/abs/arXiv:1906.10980} {arXiv:1906.10980 [nucl-th]}
\BibitemShut {NoStop}%
\bibitem [{\citenamefont {Abdurashitov}\ \emph {et~al.}(2006)\citenamefont
{Abdurashitov} \emph {et~al.}}]{Abdurashitov:2005tb}%
\BibitemOpen
\bibfield {author} {\bibinfo {author} {\bibfnamefont {J.~N.}\ \bibnamefont
{Abdurashitov}} \emph {et~al.} (\bibinfo {collaboration} {SAGE}),\
}\href@noop {} {\bibfield {journal} {\bibinfo {journal} {Phys. Rev.}\
}\textbf {\bibinfo {volume} {C73}},\ \bibinfo {pages} {045805} (\bibinfo
{year} {2006})},\ \Eprint {http://arxiv.org/abs/nucl-ex/0512041}
{nucl-ex/0512041} \BibitemShut {NoStop}%
\bibitem [{\citenamefont {Laveder}(2007)}]{Laveder:2007zz}%
\BibitemOpen
\bibfield {author} {\bibinfo {author} {\bibfnamefont {M.}~\bibnamefont
{Laveder}},\ }\href {\doibase 10.1016/j.nuclphysbps.2007.02.037} {\bibfield
{journal} {\bibinfo {journal} {Nucl. Phys. Proc. Suppl.}\ }\textbf {\bibinfo
{volume} {168}},\ \bibinfo {pages} {344} (\bibinfo {year}
{2007})}\BibitemShut {NoStop}%
\bibitem [{\citenamefont {Giunti}\ and\ \citenamefont
{Laveder}(2007)}]{Giunti:2006bj}%
\BibitemOpen
\bibfield {author} {\bibinfo {author} {\bibfnamefont {C.}~\bibnamefont
{Giunti}}\ and\ \bibinfo {author} {\bibfnamefont {M.}~\bibnamefont
{Laveder}},\ }\href@noop {} {\bibfield {journal} {\bibinfo {journal} {Mod.
Phys. Lett.}\ }\textbf {\bibinfo {volume} {A22}},\ \bibinfo {pages} {2499}
(\bibinfo {year} {2007})},\ \Eprint {http://arxiv.org/abs/hep-ph/0610352}
{hep-ph/0610352} \BibitemShut {NoStop}%
\bibitem [{\citenamefont {Acero}\ \emph {et~al.}(2008)\citenamefont {Acero},
\citenamefont {Giunti},\ and\ \citenamefont {Laveder}}]{Acero:2007su}%
\BibitemOpen
\bibfield {author} {\bibinfo {author} {\bibfnamefont {M.~A.}\ \bibnamefont
{Acero}}, \bibinfo {author} {\bibfnamefont {C.}~\bibnamefont {Giunti}}, \
and\ \bibinfo {author} {\bibfnamefont {M.}~\bibnamefont {Laveder}},\
}\href@noop {} {\bibfield {journal} {\bibinfo {journal} {Phys. Rev.}\
}\textbf {\bibinfo {volume} {D78}},\ \bibinfo {pages} {073009} (\bibinfo
{year} {2008})},\ \Eprint {http://arxiv.org/abs/arXiv:0711.4222}
{arXiv:0711.4222 [hep-ph]} \BibitemShut {NoStop}%
\bibitem [{\citenamefont {Giunti}\ and\ \citenamefont
{Laveder}(2009)}]{Giunti:2009zz}%
\BibitemOpen
\bibfield {author} {\bibinfo {author} {\bibfnamefont {C.}~\bibnamefont
{Giunti}}\ and\ \bibinfo {author} {\bibfnamefont {M.}~\bibnamefont
{Laveder}},\ }\href {\doibase 10.1103/PhysRevD.80.013005} {\bibfield
{journal} {\bibinfo {journal} {Phys. Rev.}\ }\textbf {\bibinfo {volume}
{D80}},\ \bibinfo {pages} {013005} (\bibinfo {year} {2009})},\ \Eprint
{http://arxiv.org/abs/arXiv:0902.1992} {arXiv:0902.1992 [hep-ph]}
\BibitemShut {NoStop}%
\bibitem [{\citenamefont {Giunti}\ and\ \citenamefont
{Laveder}(2011)}]{Giunti:2010zu}%
\BibitemOpen
\bibfield {author} {\bibinfo {author} {\bibfnamefont {C.}~\bibnamefont
{Giunti}}\ and\ \bibinfo {author} {\bibfnamefont {M.}~\bibnamefont
{Laveder}},\ }\href@noop {} {\bibfield {journal} {\bibinfo {journal} {Phys.
Rev.}\ }\textbf {\bibinfo {volume} {C83}},\ \bibinfo {pages} {065504}
(\bibinfo {year} {2011})},\ \Eprint {http://arxiv.org/abs/arXiv:1006.3244}
{arXiv:1006.3244 [hep-ph]} \BibitemShut {NoStop}%
\bibitem [{\citenamefont {Anselmann}\ \emph {et~al.}(1995)\citenamefont
{Anselmann} \emph {et~al.}}]{Anselmann:1994ar}%
\BibitemOpen
\bibfield {author} {\bibinfo {author} {\bibfnamefont {P.}~\bibnamefont
{Anselmann}} \emph {et~al.} (\bibinfo {collaboration} {GALLEX}),\ }\href@noop
{} {\bibfield {journal} {\bibinfo {journal} {Phys. Lett.}\ }\textbf
{\bibinfo {volume} {B342}},\ \bibinfo {pages} {440} (\bibinfo {year}
{1995})}\BibitemShut {NoStop}%
\bibitem [{\citenamefont {Hampel}\ \emph {et~al.}(1998)\citenamefont {Hampel}
\emph {et~al.}}]{Hampel:1997fc}%
\BibitemOpen
\bibfield {author} {\bibinfo {author} {\bibfnamefont {W.}~\bibnamefont
{Hampel}} \emph {et~al.} (\bibinfo {collaboration} {GALLEX}),\ }\href@noop {}
{\bibfield {journal} {\bibinfo {journal} {Phys. Lett.}\ }\textbf {\bibinfo
{volume} {B420}},\ \bibinfo {pages} {114} (\bibinfo {year}
{1998})}\BibitemShut {NoStop}%
\bibitem [{\citenamefont {Kaether}\ \emph {et~al.}(2010)\citenamefont
{Kaether}, \citenamefont {Hampel}, \citenamefont {Heusser}, \citenamefont
{Kiko},\ and\ \citenamefont {Kirsten}}]{Kaether:2010ag}%
\BibitemOpen
\bibfield {author} {\bibinfo {author} {\bibfnamefont {F.}~\bibnamefont
{Kaether}}, \bibinfo {author} {\bibfnamefont {W.}~\bibnamefont {Hampel}},
\bibinfo {author} {\bibfnamefont {G.}~\bibnamefont {Heusser}}, \bibinfo
{author} {\bibfnamefont {J.}~\bibnamefont {Kiko}}, \ and\ \bibinfo {author}
{\bibfnamefont {T.}~\bibnamefont {Kirsten}},\ }\href@noop {} {\bibfield
{journal} {\bibinfo {journal} {Phys. Lett.}\ }\textbf {\bibinfo {volume}
{B685}},\ \bibinfo {pages} {47} (\bibinfo {year} {2010})},\ \Eprint
{http://arxiv.org/abs/arXiv:1001.2731} {arXiv:1001.2731 [hep-ex]}
\BibitemShut {NoStop}%
\bibitem [{\citenamefont {Abdurashitov}\ \emph {et~al.}(1996)\citenamefont
{Abdurashitov} \emph {et~al.}}]{Abdurashitov:1996dp}%
\BibitemOpen
\bibfield {author} {\bibinfo {author} {\bibfnamefont {J.~N.}\ \bibnamefont
{Abdurashitov}} \emph {et~al.} (\bibinfo {collaboration} {SAGE}),\
}\href@noop {} {\bibfield {journal} {\bibinfo {journal} {Phys. Rev. Lett.}\
}\textbf {\bibinfo {volume} {77}},\ \bibinfo {pages} {4708} (\bibinfo {year}
{1996})}\BibitemShut {NoStop}%
\bibitem [{\citenamefont {Abdurashitov}\ \emph {et~al.}(1999)\citenamefont
{Abdurashitov} \emph {et~al.}}]{Abdurashitov:1998ne}%
\BibitemOpen
\bibfield {author} {\bibinfo {author} {\bibfnamefont {J.~N.}\ \bibnamefont
{Abdurashitov}} \emph {et~al.} (\bibinfo {collaboration} {SAGE}),\
}\href@noop {} {\bibfield {journal} {\bibinfo {journal} {Phys. Rev.}\
}\textbf {\bibinfo {volume} {C59}},\ \bibinfo {pages} {2246} (\bibinfo {year}
{1999})},\ \Eprint {http://arxiv.org/abs/hep-ph/9803418} {hep-ph/9803418}
\BibitemShut {NoStop}%
\bibitem [{\citenamefont {Abdurashitov}\ \emph {et~al.}(2009)\citenamefont
{Abdurashitov} \emph {et~al.}}]{Abdurashitov:2009tn}%
\BibitemOpen
\bibfield {author} {\bibinfo {author} {\bibfnamefont {J.~N.}\ \bibnamefont
{Abdurashitov}} \emph {et~al.} (\bibinfo {collaboration} {SAGE}),\
}\href@noop {} {\bibfield {journal} {\bibinfo {journal} {Phys. Rev.}\
}\textbf {\bibinfo {volume} {C80}},\ \bibinfo {pages} {015807} (\bibinfo
{year} {2009})},\ \Eprint {http://arxiv.org/abs/arXiv:0901.2200}
{arXiv:0901.2200 [nucl-ex]} \BibitemShut {NoStop}%
\bibitem [{\citenamefont {Krofcheck}\ \emph {et~al.}(1985)\citenamefont
{Krofcheck} \emph {et~al.}}]{Krofcheck:1985fg}%
\BibitemOpen
\bibfield {author} {\bibinfo {author} {\bibfnamefont {D.}~\bibnamefont
{Krofcheck}} \emph {et~al.},\ }\href {\doibase 10.1103/PhysRevLett.55.1051}
{\bibfield {journal} {\bibinfo {journal} {Phys. Rev. Lett.}\ }\textbf
{\bibinfo {volume} {55}},\ \bibinfo {pages} {1051} (\bibinfo {year}
{1985})}\BibitemShut {NoStop}%
\end{thebibliography}

\end{document}